\documentclass[final,5p,times,twocolumn]{elsarticle}

\usepackage{amssymb}
\usepackage{amsthm}
\usepackage{amsmath}
\usepackage[english]{babel}
\usepackage{xcolor}
\usepackage[style=base]{subcaption}
\usepackage{float}
\usepackage{textcomp}
\usepackage{gensymb}
\usepackage{url}
\usepackage{hyperref}
\usepackage{graphicx}
\usepackage[T1]{fontenc}
\usepackage{pifont} 

\journal{Journal of Magnetic Resonance}

\begin{document}

\begin{frontmatter}



\title{A Ramsey apparatus for proton spins in flowing water}


\author[lhep]{Ivo~Schulthess}\corref{cor1}
\ead{ivo.schulthess@lhep.unibe.ch}

\author[lhep]{Anastasio~Fratangelo}

\author[lin]{Patrick~Hautle}

\author[lhep]{Philipp~Heil}

\author[lhep]{Gjon Markaj}

\author[lhep]{Marc~Persoz}

\author[lhep]{Ciro~Pistillo}

\author[lhep]{Jacob~Thorne}

\author[lhep]{Florian~M.~Piegsa}\corref{cor1}
\ead{florian.piegsa@lhep.unibe.ch}

\cortext[cor1]{Corresponding author}

\affiliation[lhep]{organization={Laboratory for High Energy Physics, Albert Einstein Center for Fundamental Physics, University of Bern},
            addressline={Sidlerstrasse~5}, 
            city={3012~Bern},
            country={Switzerland}}

\affiliation[lin]{organization={Laboratory for Neutron and Muon Instrumentation (LIN), Paul Scherrer Institut}, 
            addressline={Forschungsstrasse~111}, 
            city={5232~Villigen~PSI}, 
            country={Switzerland}}

\begin{abstract}
We present an apparatus that applies Ramsey's method of separated oscillatory fields to proton spins in water molecules. The setup consists of a water circuit, a spin polarizer, a magnetically shielded interaction region with various radio frequency elements, and a nuclear magnetic resonance system to measure the spin polarization. We show that this apparatus can be used for Rabi resonance measurements and to investigate magnetic and pseudomagnetic field effects in Ramsey-type precision measurements with a sensitivity below 100~pT. 
\end{abstract}




\end{frontmatter}


\section{Introduction}\label{sec:intro}

The nuclear magnetic resonance method of Rabi~\cite{rabi_molecular_1939, kellogg_magnetic_1939} and Ramsey’s technique of separated oscillatory fields~\cite{ramsey_new_1949,ramsey_molecular_1950,ramsey_neutron_1986} have been applied very successfully in a variety of different scientific experiments. They apply constant and time varying magnetic fields to manipulate the spins of probe particles. 

Ramsey's technique allows to determine the Larmor precession frequency of the spin in a magnetic field $B_0$. In a first step, the spin polarized particles are flipped by an oscillating field $B_1$ into the plane orthogonal to $B_0$. Then, they can precess for a certain time until they are flipped again by a second oscillating $B_1$ field. Usually, the frequency of the oscillating fields is scanned close to the resonance while the phases of the two signals are locked. This results in an interference pattern of the spin polarization in the frequency domain. Ramsey's technique can be applied to precisely measure changes in magnetic and pseudo-magnetic fields. It is used in atomic clocks~\cite{essen_atomic_1955,wynands_atomic_2005}, to measure the Newtonian gravitational constant~\cite{rosi_precision_2014}, to search for the neutron electric dipole moment~\cite{abel_measurement_2020,piegsa_new_2013,chupp_electric_2019}, to search for dark matter~\cite{abel_search_2017,schulthess_new_2022}, new particles and interactions~\cite{piegsa_limits_2012}, and others. It was also applied in the measurement of the neutron magnetic moment~\cite{greene_measurement_1979}. In the latter experiment, the technique served to compare resonance frequencies of free neutrons and protons in water passing through one apparatus. The application of resonance techniques with flowing water had been previously demonstrated by Sherman~\cite{sherman_nuclear_1954}. 

We present an experimental apparatus with a concept similar to the one used in the measurement of the neutron magnetic moment. Ultimately, with our setup, we aim to perform laboratory searches for new exotic long-range interactions beyond the standard model of particle physics~\cite{dobrescu_spin-dependent_2006,piegsa_limits_2012,schulthess_search_2022}. Such a coupling has so far never been investigated using protons as probe particles. The originality of the apparatus makes it an excellent tool for systematically testing pulse sequences and investigating various magnetic resonance effects, for instance, the Bloch-Siegert shift~\cite{bloch_magnetic_1940} or dressed spin states~\cite{cohen-tannoudji_absorption_1969,muskat_dressed_1987}. It can also serve as a co-magnetometer in other precision physics experiments.

\section{Apparatus}\label{sec:apparatus}

\begin{figure}[!t]
    \centering
    \includegraphics[width=0.48\textwidth]{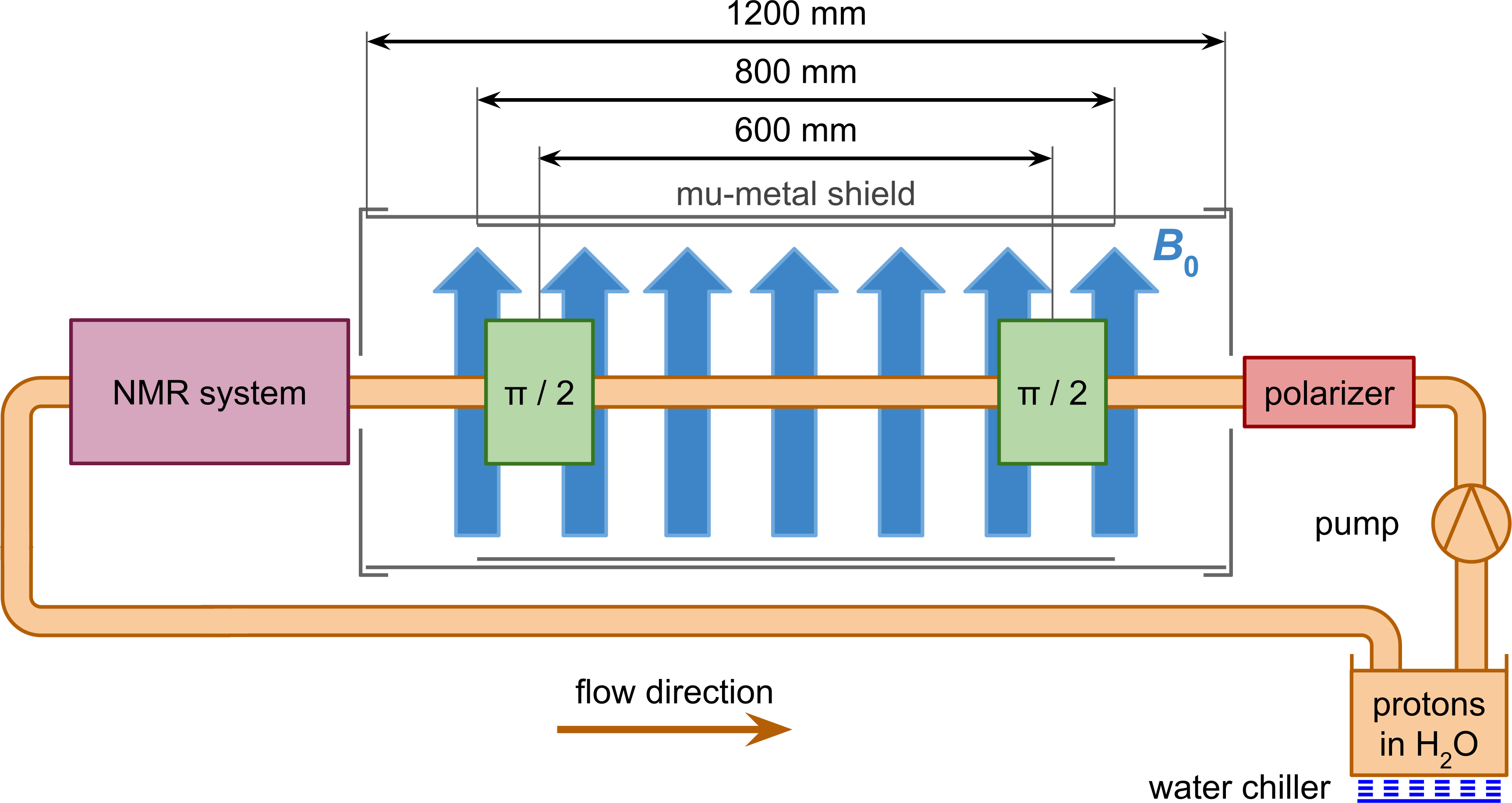}
    \caption{Schematic of the experimental setup where the protons in water (H$_2$O) are pumped from the water reservoir of the chiller. They are first polarized in a polarizer (red) and then enter the interaction region surrounded by a double-layer mu-metal shield. Two spin-flip coils are shown in green and the magnetic field direction is indicated in blue. The spin polarization is analyzed using a nuclear magnetic resonance (NMR) system (purple). The schematic is not to scale. }
    \label{fig:experimentalSchematic}
\end{figure}

\begin{figure}[!t]
    \centering
    \includegraphics[width=0.48\textwidth]{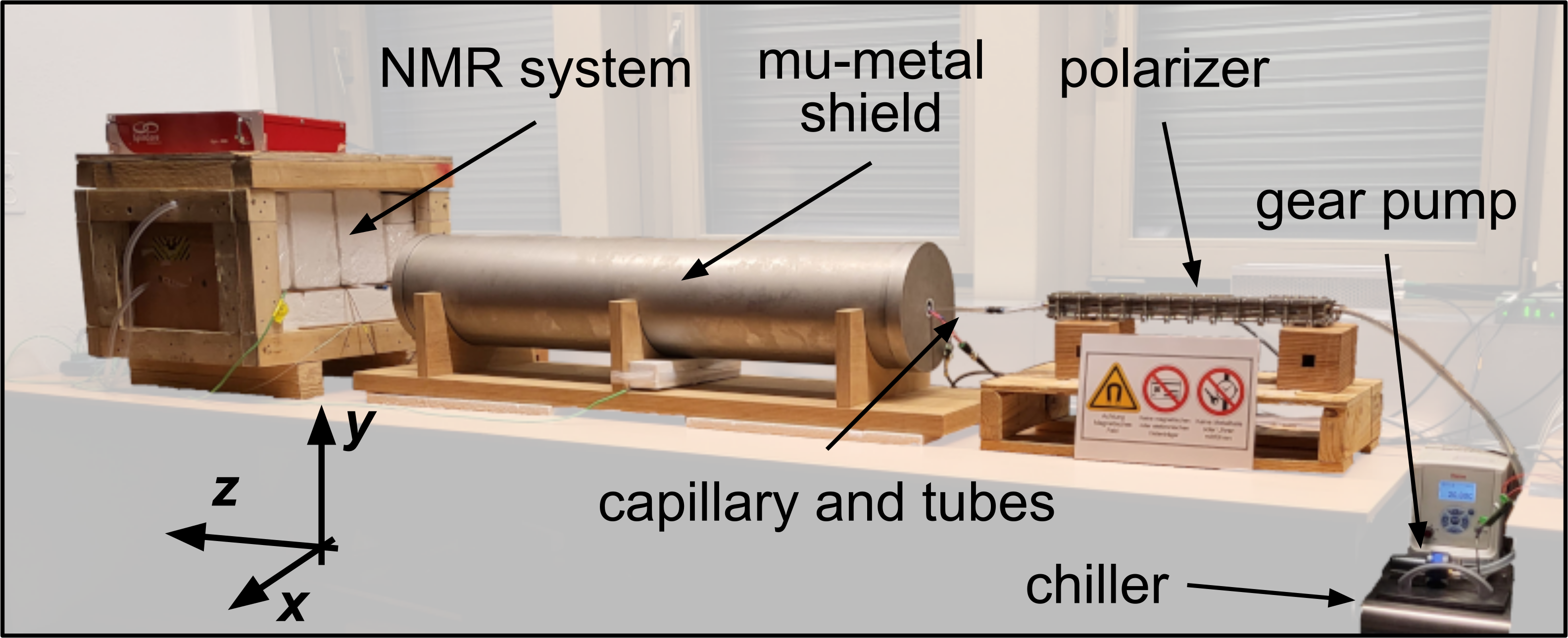}
    \caption{Photo of the experimental setup. The gear pump is mounted on the water chiller and pumps the water from right to left through the setup. First, the water passes through the polarizer. It then flows in a glass capillary through the interaction region surrounded by the mu-metal shield. Finally, the spin polarization is measured with an NMR system. The NMR magnet is temperature stabilized and thermally insulated.}
    \label{fig:experimentalSetup}
\end{figure}

A schematic of the experiment is depicted in Fig.~\ref{fig:experimentalSchematic} and a photo of the full tabletop apparatus is shown in Fig.~\ref{fig:experimentalSetup}. The total length is about 3~meters. The water is circulated through the system using a gear pump. First, the water passes a polarizer to create a sizable spin polarization of the protons. It then flows through the interaction region, which is magnetically shielded to the surrounding by mu-metal. In that region, the spins interact with the magnetic field $B_0$ and can be manipulated with spin-flip coils. There are additional temperature and magnetic field sensors. Finally, the spin polarization is measured and analyzed employing nuclear magnetic resonance (NMR) techniques. No guiding fields for the proton spins are required between the elements since their fringe fields are sufficient. 

\subsection{Water Circuit}\label{sec:waterCircuit}

\begin{figure}[!t]
    \centering
    \includegraphics[width=0.48\textwidth]{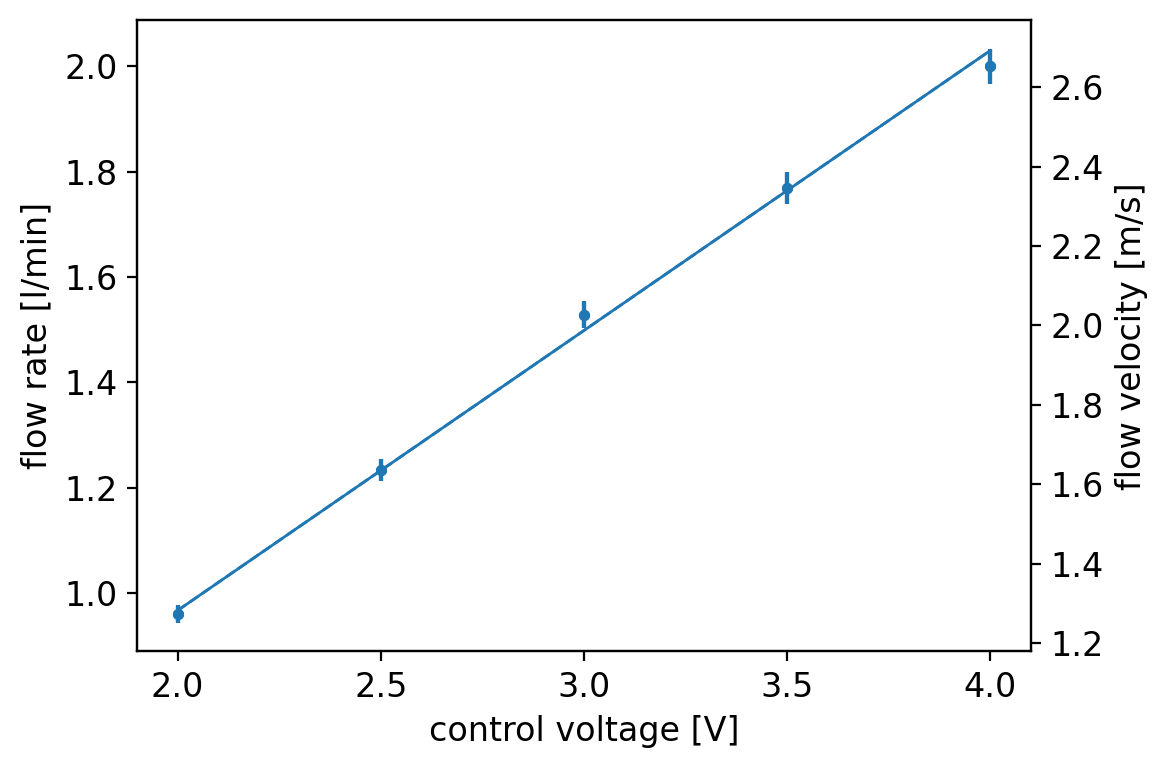}
    \caption{Flow rate as a function of the control voltage applied to the pump. The vertical axis on the right side shows the average velocity of the water through the glass capillary in the interaction region with an inner diameter of 4~mm. The solid line is a linear fit through the data. }
    \label{fig:pumpFlowRate}
\end{figure}

To perform a spin precession experiment, the demineralized water that contains the hydrogen protons is circulating in a water circuit. We use a rigid glass capillary with an inner diameter of ${d=4~\mathrm{mm}}$ and a length of 1500~mm to guide the water through the interaction region. To connect the other elements we use plastic tubes (PU, PVC, and PTFE) of various diameters. We use a gear pump MGD2000F~\cite{tcs_micropumps_ltd_mgd2000_2021} with flow and pressure ratings of 2.3~l/min and 6~bar, respectively. This is suitable to transport the water through the setup within a few seconds to maintain an usable degree of polarization and overcome all pressure losses of the system. The power of the pump can be adjusted with an external control voltage. The measured flow rate versus the applied voltage is shown in Fig.~\ref{fig:pumpFlowRate}. It also shows the corresponding average velocity through the glass capillary in the interaction region since this is an interesting property for many resonance measurements. The flow rate and velocity is linear in the measured range and a linear fit led to the values of $(0.53 \pm 0.02)$~l/(min$\cdot$V) and $(0.71 \pm 0.02)$~m/(s$\cdot$V), respectively. 

The flow velocity was set to $v=2.35$~m/s in all measurements if not stated otherwise. The velocity was optimized for highest signal visibility while operating the pump and the water system within the safety margins of the pressure ratings. The pump is mounted on a chiller, which is a temperature stabilized bath circulator Thermo Scientific ARCTIC A10-SC150~\cite{thermo_fisher_scientific_thermo_2015}. The pump uses the water from the chiller's reservoir that is kept at a temperature of 20~$\degree$C to minimize systematic effects like the temperature dependence of the proton resonance frequency~\cite{petley_temperature_1984,cho_nmr_2002}.

\subsection{Polarizer}\label{sec:polarizer}

\begin{figure}[!t]
  \centering
  \begin{subfigure}[b]{0.48\textwidth}
    \includegraphics[width=\textwidth]{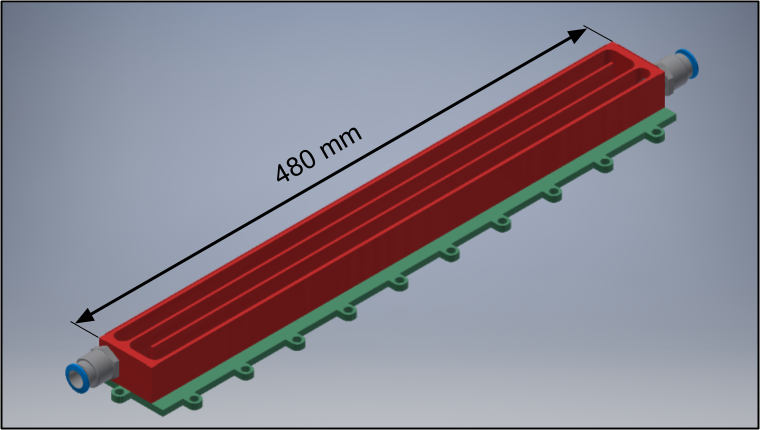}
    \caption{~}
    \label{fig:polarizerCAD}
  \end{subfigure}
  \hfill
  \begin{subfigure}[b]{0.48\textwidth}
    \includegraphics[width=\textwidth]{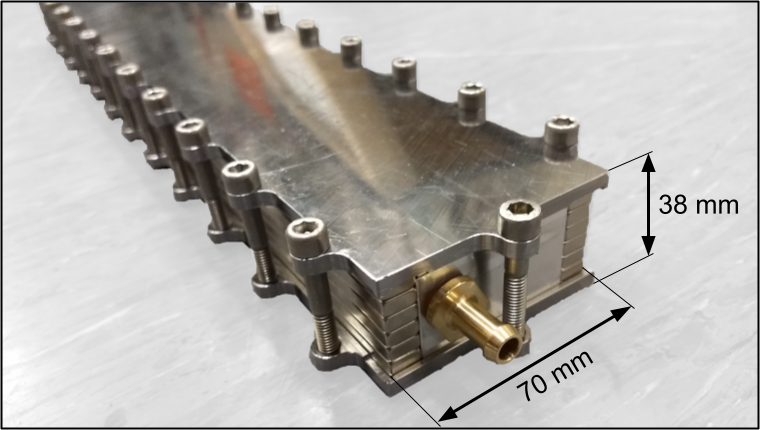}
    \caption{~}
    \label{fig:polarizerPhoto}
  \end{subfigure}
  \caption{(a) CAD model of the polarizer that shows the meandering groove on the inside of the aluminum body (red). The steel base plate (green) below the body is also shown. The permanent magnets on the two sides of the body are not shown. (b) Photo of the polarizer with the water inlet in front. The two steel plates are mounted together with screws to hold the magnets left and right of the aluminum body in place. }
  \label{fig:polarizer}
\end{figure}

The spin of the particles is polarized by guiding the water through a strong external magnetic field which is applied by a polarizer. It has an aluminum body where the water flows through a meandering groove. Tube connections for the water inlet and outlet are installed on both ends of the body. The outer cross-section of the body is $30 \times 50$~mm$^2$ and it has a length of 480~mm. Its total water volume is 420~ml. Neodymium permanent magnets are stacked on both sides of the aluminum body. For small volumes, this approach is much simpler and cheaper than an electromagnet as it does not require high currents or rely on water cooling. The magnets used are Q-40-10-05-N with grade N42~\cite{supermagnete_datenblatt_2018}. They have a size of $40 \times 10 \times 5$~mm$^3$. To create a homogeneous magnetic field over the full polarizer volume, steel plates with a thickness of 4~mm cover the body on both sides. The plates are also used to hold the body and all the magnets in place. This design provides a magnetic field in the interior of roughly $190$~mT. The field was estimated with a simulation using the \textit{Finite Element Method Magnetics} software~\cite{david_meeker_finite_2019} and measured using a Hall-probe~\cite{magnet-physik_dr_steingroever_gmbh_usb_2020}. The CAD model of the polarizer's interior and a photo of the actual polarizer are shown in Fig.~\ref{fig:polarizerCAD} and Fig.~\ref{fig:polarizerPhoto}, respectively. 

The polarization of the proton spins in an external magnetic field as a function of the exposure time follows an exponential law~\cite{bloch_nuclear_1946}
\begin{equation}\label{eq:polarization}
    P(t) = P_0 \times (1 - e^{-t/T_1}) \ ,
\end{equation}
where $P_0 = \tanh \left( \frac{\gamma_p \hbar B_p}{2 k_B T} \right)$ is the thermal equilibrium polarization of the proton spins that depends only on the magnetic field strength of the polarizer $B_p$ and the water temperature $T$~\cite{slichter_principles_1978}. Here, $\gamma_p$ is the gyromagnetic ratio of the proton, $\hbar$ the reduced Planck constant, and $k_B$ the Boltzmann constant. With the given magnetic field and a temperature of $T = 20~\degree$C, this yields a polarization of $P_0 \approx 7 \times 10^{-7}$. ${T_1 = (2.35\pm0.02)~\mathrm{s}}$ is the longitudinal or spin-lattice relaxation time constant which was determined in an auxiliary measurement using an inversion recovery pulse sequence~\cite{hahn_accurate_1949}. To achieve a polarization of $P(t) > 0.99 \cdot P_0$ the water has to spend $5 \cdot T_1 \approx 12$~s within the polarizer. With a volume of 420~ml, the water spends more than five time constants in the polarizer, even for the highest measured flow rates. There exists a second type of relaxation called transversal or spin-spin relaxation with a corresponding time constant $T_2$. The latter describes the dephasing time of the spins in the plane perpendicular to the external magnetic field $B_0$ due to their mutual interaction. We used a CPMG pulse sequence~\cite{carr_effects_1954,meiboom_modified_1958} in a separate measurement and determined $T_2 = (1.67 \pm 0.02)$~s. The transversal relaxation time is not critical for our measurements.

\subsection{Interaction Region}\label{sec:interactionRegion}

\begin{figure}[!t]
  \centering
  \begin{subfigure}[b]{0.48\textwidth}
    \includegraphics[width=\textwidth]{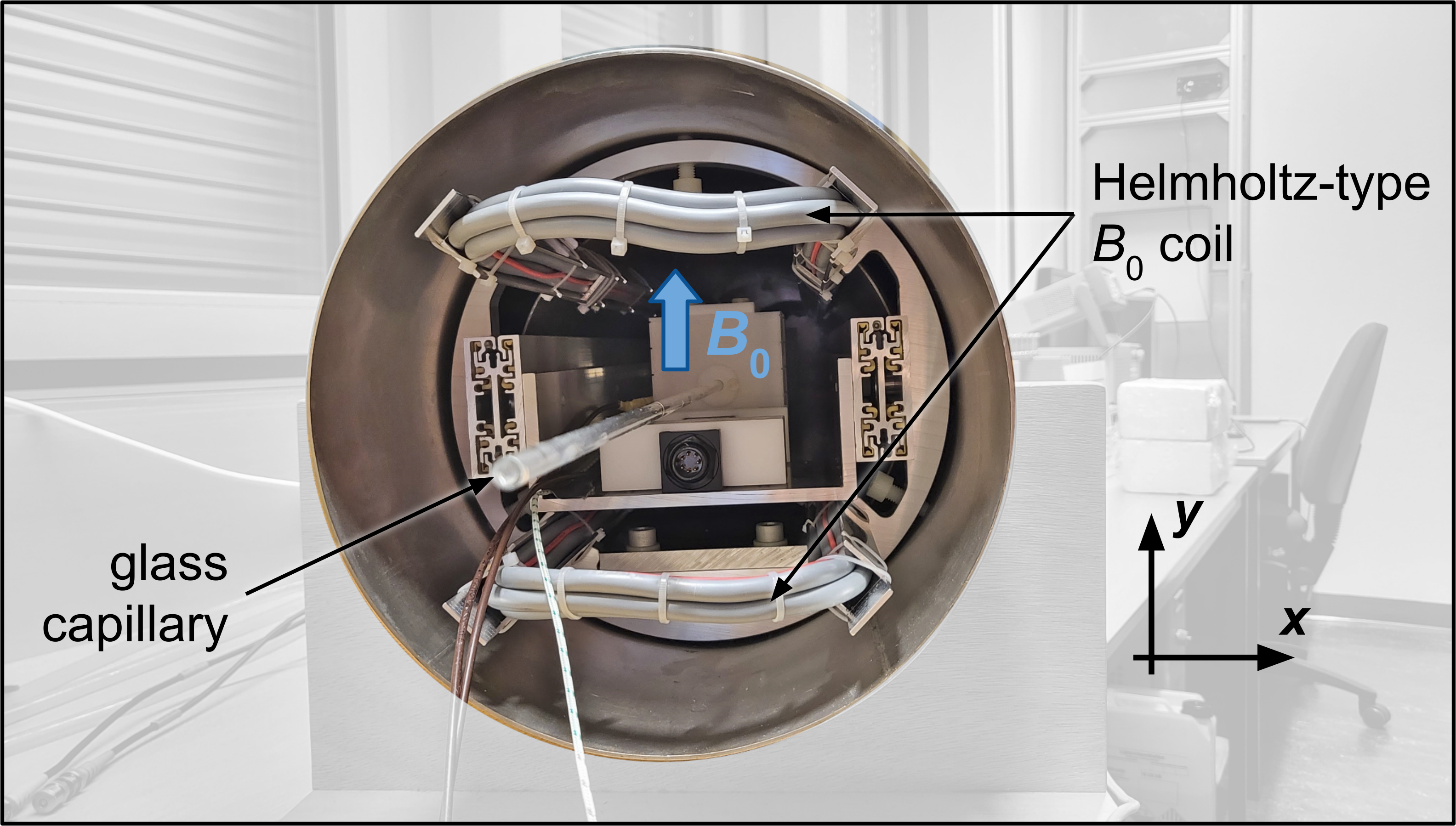}
    \caption{~}
    \label{fig:muMetalInterior}
  \end{subfigure}
  \hfill
  \begin{subfigure}[b]{0.48\textwidth}
    \includegraphics[width=\textwidth]{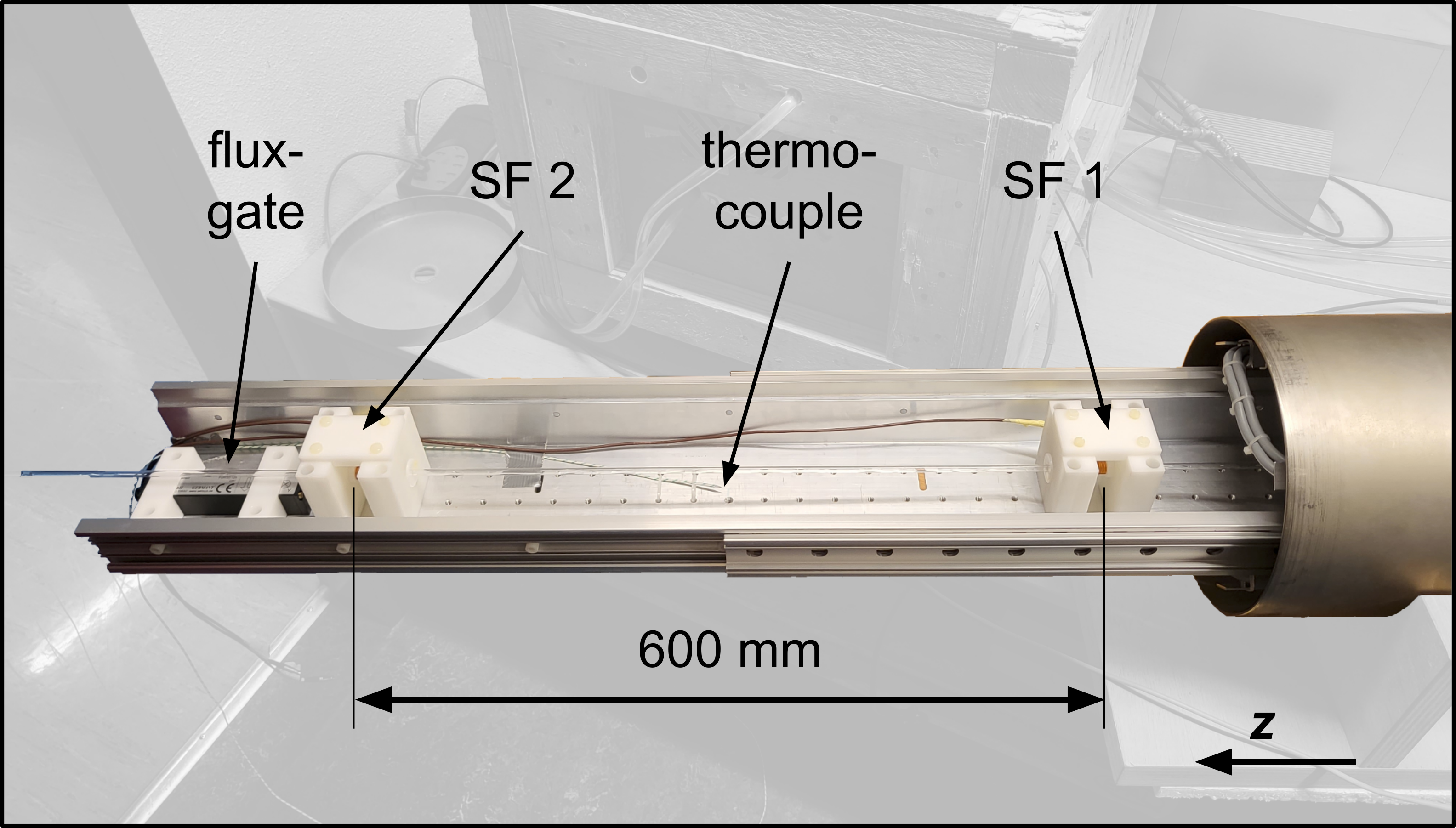}
    \caption{~}
    \label{fig:interactionZone}
  \end{subfigure}
  \caption{(a) Photo of the view into the mu-metal. The inner mu-metal layer is hidden behind an aluminum ring that holds the telescopic-rails. The water flows through the glass capillary in the center. The ends of the rectangular-shaped Helmholtz coil are also visible. (b) Photo of the interaction region slid out of the mu-metal. Water flows along the $z$-axis from right to left. The proton spins can be manipulated with the spin-flip coils SF~1 and SF~2. A non-magnetic thermocouple type-E measures the temperature inside and a fluxgate sensor behind SF~2 records the magnetic field in all three spatial directions.  }
  \label{fig:interactionRegion}
\end{figure}

The interaction region is where we apply the Rabi and Ramsey technique to the proton spins. It is surrounded by a passive magnetic shield made of two cylindrical mu-metal layers as shown in Fig.~\ref{fig:muMetalInterior}. The outer layer has an inner diameter of 235~mm and a length of 1200~mm. The inner layer is concentric within the outer layer, has an inner diameter of 195~mm, and a length of 800~mm. Both layers have a thickness of 2~mm. End caps for the outer layer are available but usually not installed to allow for easy access from the outside. The main magnetic field $B_0$ is aligned along the vertical $y$-direction. It is created using a rectangular-shaped Helmholtz-type coil with 20~windings that is visible in Fig.~\ref{fig:muMetalInterior}. The coil has a width and separation of 128~mm and a length of 1200~mm. It is centered with respect to the mu-metal shield and is connected to a Keysight B2962A Low Noise Power Source~\cite{keysight_technologies_keysight_2020}. We measured a field constant of the coil of 232~$\mu$T/A. Since the interior of the mu-metal shield is difficult to access, an aluminum U-profile with a grid of threads is mounted to non-magnetic telescopic-rails that are fixed to the mu-metal tube. It can be slid out of the mu-metal to change the arrangement of spin-flip coils, fluxgate sensors, and temperature sensors. 

The standard arrangement is shown in Fig.~\ref{fig:interactionZone}. It corresponds to a Ramsey setup that consists of two spin-flip coils with a center-to-center separation of 600~mm. The spin-flip coils are solenoids with 16~windings, and their axis is aligned with the water pipe. The coils are made of copper with a wire diameter of 0.8~mm. They have an inner diameter of 10~mm, and a length of 15~mm. They are held by a POM holder block that allows the mounting of a trimming coil to adjust the local magnetic field around each spin-flip coil. Optionally, a third spin-flip coil can be mounted in the center to perform spin-echo measurements. 

\begin{figure}[!t]
    \centering
    \includegraphics[width=0.48\textwidth]{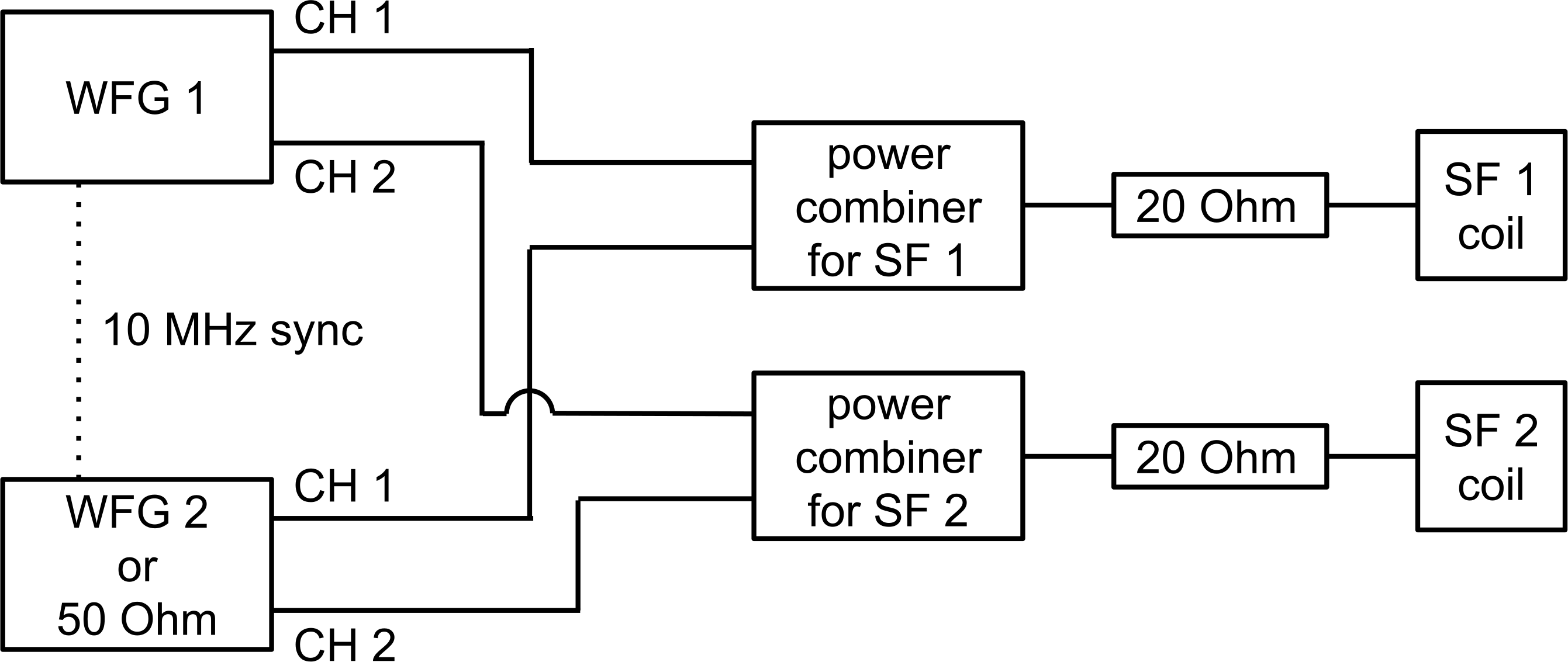}
    \caption{Connection diagram of the spin-flip coils. The waveform generator (WFG~1) produces the oscillating currents. These signals can be combined with the oscillating signals of a second waveform generator (WFG~2) or terminated with $50~\Omega$ resistors if not used. The outputs of the combiners are connected to the spin-flip coils (SF~1 and SF~2) via $20~\Omega$ resistors. }
    \label{fig:sfConnectionDiagram}
\end{figure}

To manipulate the proton spins, the spin-flip coils have to be driven with oscillating currents close to the Larmor resonance frequency. A connection diagram is shown in Fig.~\ref{fig:sfConnectionDiagram}. The oscillating currents are provided by two Keysight waveform generators 33622A~\cite{keysight_technologies_keysight_2021}. For each spin-flip coil an output channel is connected to a Mini-Circuits power combiner ZFRSC-2050+~\cite{mini-circuits_coaxial_2011}. The second input of the combiner can be used to induce signals at different frequencies to test effects like the Bloch-Siegert shift~\cite{bloch_magnetic_1940} or the dressed spin states~\cite{cohen-tannoudji_absorption_1969}. These second inputs are terminated with $50~\Omega$ if they are not used. The output of each combiner is connected to a spin-flip coil via a $20~\Omega$ resistor to reduce the frequency dependence of the impedance. 

A non-magnetic thermocouple type-E measures the temperature in the center of the interaction region close to the glass capillary.  Outside of the magnetic shield, two thermocouples of type-K measure the temperature of the room below the mu-metal and of the NMR magnet. They are read out with a rate of 2~Hz and a precision better then 0.025~$\degree$C using a Picotech data logger TC-08~\cite{pico_technology_tc-08_2021}.  

The magnetic field is measured using a SENSYS FGM3D/125 fluxgate sensor~\cite{sensys_gmbh_sensys_2019}. It can measure magnetic fields up to $\pm 125~\mu$T with a precision better than 150~pT in all three spatial directions. It is read out at a rate of 10~kHz using a NI~PXI-6289 analog-digital converter~\cite{national_instruments_corporation_pcipxiusb-6289_2022}. The data are averaged over 1000~samples and stored at a rate of 10~Hz. The fluxgate is located below the water tube just after the interaction region as shown in Fig~\ref{fig:interactionRegion}. It cannot be placed between the spin-flip coils since it is slightly magnetic itself and would destroy the spin coherence. The fluxgate data can be used to monitor and compensate for external magnetic field changes.

\subsection{NMR System}\label{sec:nmrSystem}

\begin{figure}[!t]
    \centering
    \includegraphics[width=0.48\textwidth]{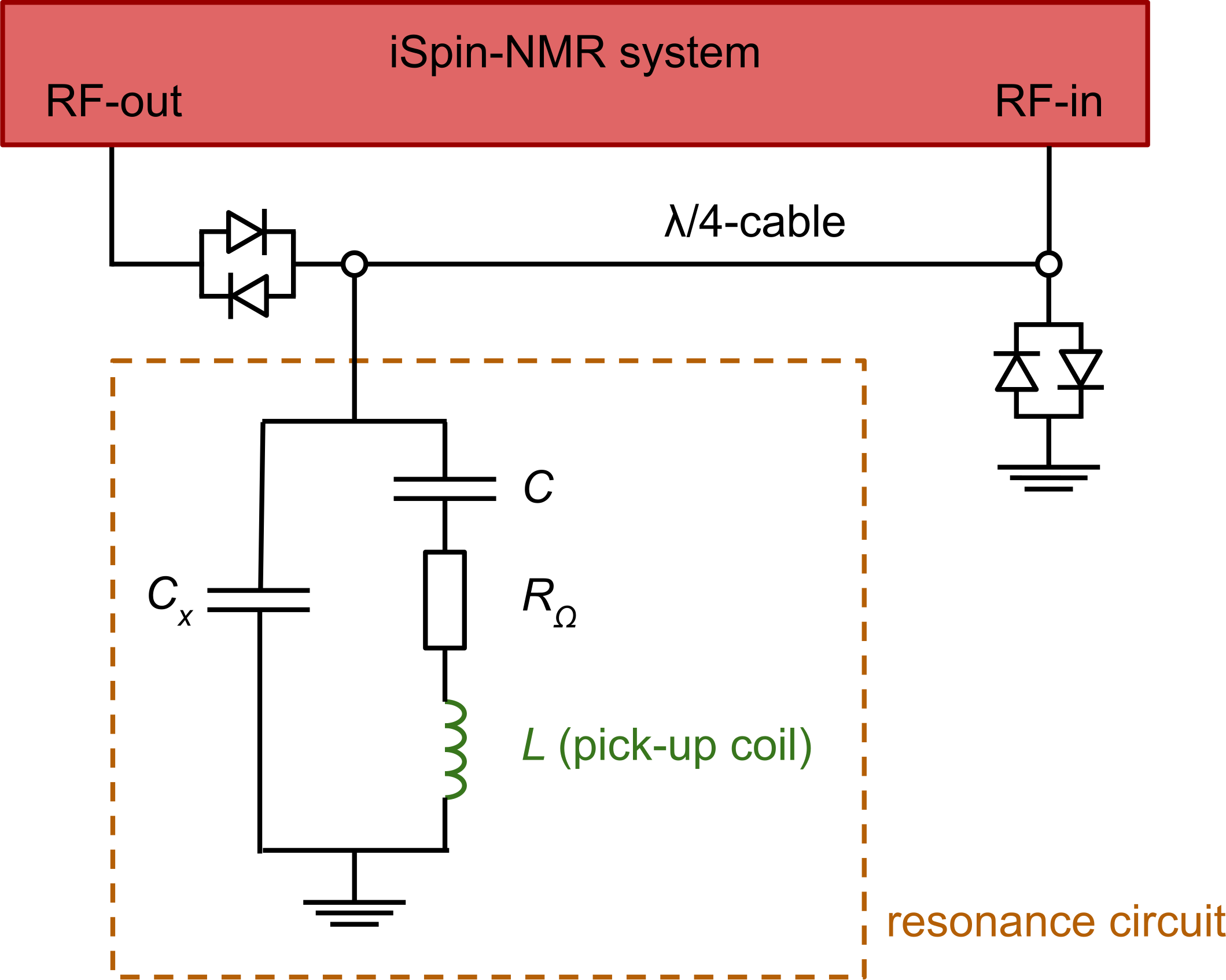}
    \caption{Connection diagram of the NMR system. The impedance of the resonance circuit of the NMR detection system has to be matched to $50~\Omega$ and tuned to the proton resonance frequency with the two capacitors $C$ and $C_x$. The $\lambda/4$-cable and the crossed diodes act as a passive switch that allows signals to go from the RF-out to the resonance circuit and from there to the RF-in, but not from the RF-out to the RF-in directly. }
    \label{fig:nmrCircuit}
\end{figure}

We use a commercial NMR system iSpin-NMR of SpinCore Technologies, Inc.~\cite{spincore_technologies_inc_ispin-nmr_2017} to measure the spin polarization after the interaction region. The system contains a pulse generator, a pulse amplifier, frequency filters, a preamplifier, an analog-digital converter, and further digital electronics. The system can be programmed to send arbitrary pulse sequences and measure the signal response of the NMR sample. It is able to detect voltages on the $\mu$V-level. The NMR setup uses the same coil to apply the radio-frequency (RF) pulse and to measure the precession signal of the protons. A duplexer is required to route the signal from the transmitter to the probe and from the probe to the receiver while protecting the receiver from the high-power signal of the transmitter. This is done by a passive transmit/receive switch using diodes and a quarter-wave impedance cable. A schematic of the connection diagram is shown in Fig.~\ref{fig:nmrCircuit}. 

\begin{figure}[!t]
    \centering
    \includegraphics[width=0.48\textwidth]{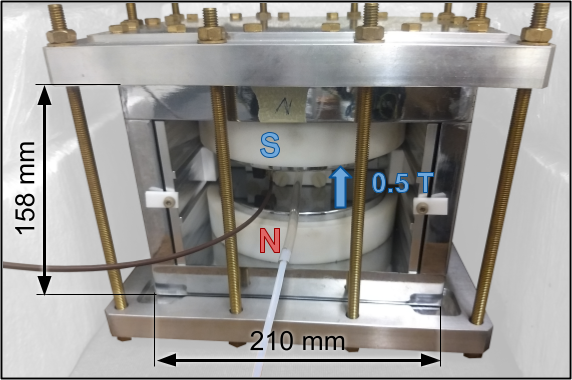}
    \caption{Photo of the NMR magnet. The two neodymium pole pieces create a magnetic field of roughly 0.5~T in the vertical direction indicated by the blue arrow. The plates for the water cooling are pressed on the magnet with brass thread rods. The holding structure for the resonance circuit and the pick-up coil made of PTFE/Teflon is visible in the center of the magnet. }
    \label{fig:nmrMagnet}
\end{figure}

To be able to transmit the power into the water sample and to measure the small precession signal, an RLC resonance circuit is required. The resonance frequency of the circuit has to be tuned to the proton resonance frequency and the impedance has to match $50~\Omega$. This can be achieved by an additional shunt-capacitor $C_x$. The inductance $L$ is given by the pick-up coil. It is made of a bare copper wire with a diameter of 0.25~mm that is isolated by a PTFE tube. The coil has 10~windings over 8~mm and a coil diameter of 4~mm. It has an bore diameter of 3~mm. The ohmic resistance $R_\Omega$ comes from the circuit cables and connections. The capacitance of the two capacitors can be calculated analytically but a final adjustment has to be done in-situ. We optimized the capacitors to $C_x=678$~pF and $C=251$~pF. 
All parts close to the NMR sample have to be made of materials that do not contain any hydrogen atoms. The precession signals of those atoms would indeed falsify the signal of the protons in the water. We used PTFE/Teflon for all parts close to the sample. 
The resonance circuit is mounted between the pole pieces of a neodymium permanent magnet. The NMR magnet has a field strength of about 0.5~T with a relative uniformity of 10$^{-4}$ for a 10~mm sample according to the manufacturer~\cite{spincore_technologies_inc_nmr_2022}. The pole pieces have a separation of 30~mm and a diameter of 140~mm. With a gyromagnetic ratio of the proton of $\gamma_p \approx 2\pi \times 42.58$~MHz/T this leads to a proton resonance frequency of 21.68~MHz~\cite{tiesinga_codata_2021}. 
We measured a relative temperature coefficient of the magnetic flux density for the NMR magnet of about -$9\times 10^{-4}$~/~$\degree$C at room temperature which is in agreement with literature values and specifications~\cite{bunting_magnetics_europe_bremag_2017,kim_temperature_1998}. Daily fluctuations of more than 1~$\degree$C lead to a change in the resonance frequency of more than 19~kHz. Since the linewidth of the proton resonance is only a few kHz, a temperature stabilization of the magnet is required. This is done via water cooling of two aluminum plates on top and below the NMR magnet. The same chiller and water bath as described in Sec.~\ref{sec:waterCircuit} are used. With this water cooling, a stability of the resonance frequency better than 1~kHz is achieved. This corresponds to temperature fluctuations of less than 0.05~$\degree$C. A photo of the NMR magnet with the cooling plates is presented in Fig.~\ref{fig:nmrMagnet}. 

Many pulse sequences exist for NMR measurements. The most basic one is a single pulse to apply a $\pi$/2-flip. Additional pulses can be applied to refocus spins (e.g. Hahn Echo~\cite{hahn_spin_1950}) or correct for other dephasing effects. Since we have a continuous flow of water through the NMR pick-up coil, only a single-pulse sequence can be applied in practice. The water in this coil gets fully replaced within 1.9~ms at a flow velocity of 2.35~m/s through the interaction region. Therefore, there is no time to apply any further pulse to the same sample and measure the spin precession signal. 
\begin{figure}[!t]
    \centering
    \includegraphics[width=0.48\textwidth]{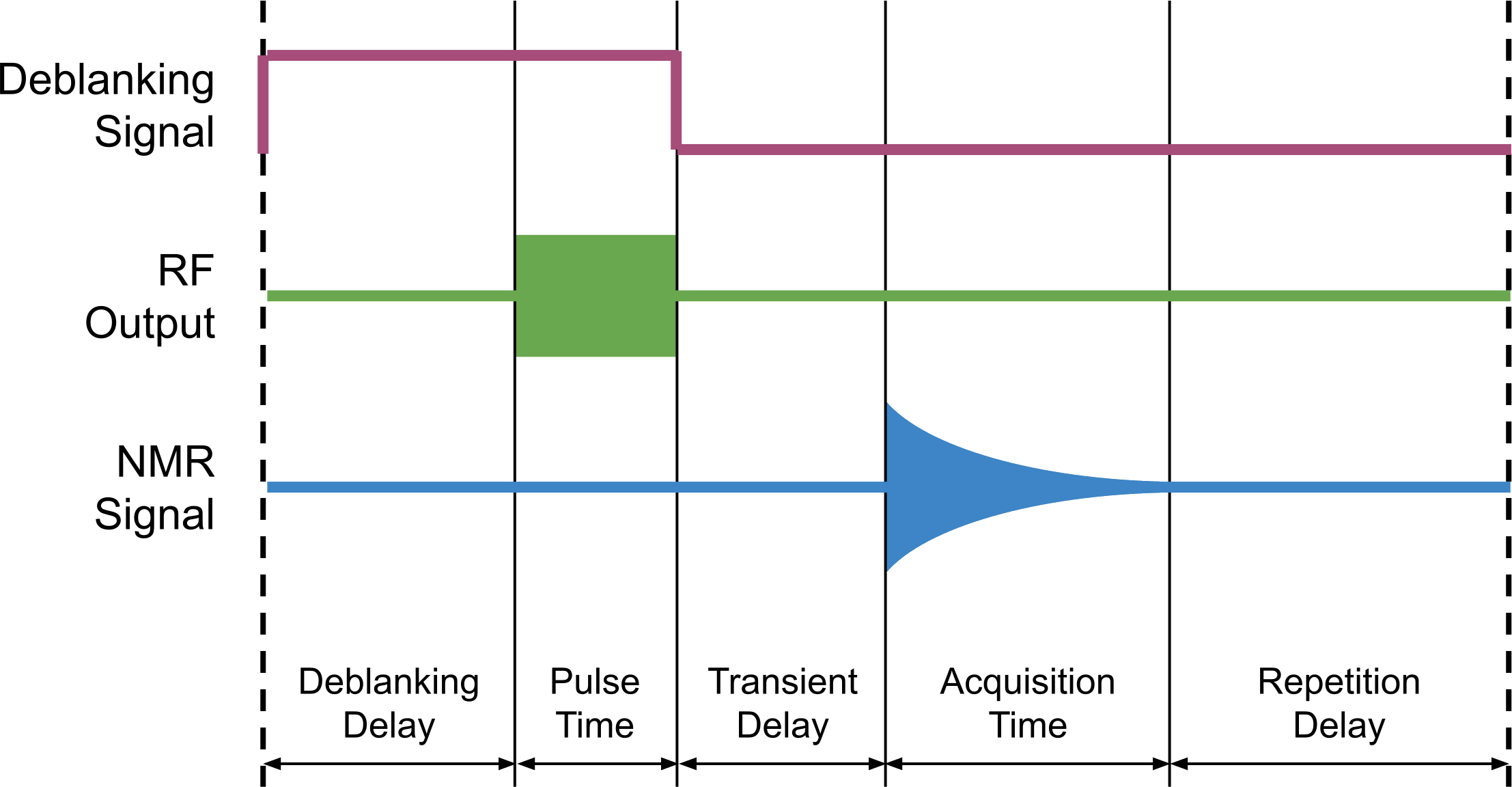}
    \caption{Diagram of the NMR data acquisition for the single-pulse sequence. Detailed description in the text. The timeline is not to scale. }
    \label{fig:nmrAcquisitionSchematic}
\end{figure}
The single-pulse sequence is diagrammed in Fig.~\ref{fig:nmrAcquisitionSchematic}. The pulse amplifier is only turned on when the deblanking signal has a logical high. The deblanking delay of 3~ms before the excitation pulse allows the amplifier to warm up. A pulse of 2.1~$\mu$s duration is then fired. The pulse time was optimized for a $\pi$/2-flip of the proton spins as this leads to the highest NMR-signal amplitude. The amplifier is then blanked (turned off) and after a short transient delay of 40~$\mu$s which allows the system to subside from any remanent pulse signal, the receiver stage starts acquiring the spin-precession signal, also called the free induction decay (FID), for 1~ms. After the data taking, the sample has to be polarized again before the measurement sequence can be repeated. In a static sample, this repetition delay is about $5 \cdot T_1 \approx 12$~s, i.e., five time constants. In our case of a continuous water flow, a repetition delay of more than 2.5~ms was chosen such that all water that underwent the NMR pulse is flushed out of the pick-up coil. Hence, the total cycle length is approximately 6.5~ms for a repetition delay of 2.5~ms. 

To improve the signal-to-noise ratio, the FID signals are averaged over many acquisitions, usually 1000~signals, before the spectral analysis is performed. Additionally, the phase of the pulse and the receiver are rotated by $\pi / 2$ for each acquisition. This way, imperfections of the two-phase detectors of the NMR system cancel out. This measurement sequence is called CYCLOPS phase cycling~\cite{keeler_understanding_2002}. 
\begin{figure}[!t]
    \centering
    \includegraphics[width=0.48\textwidth]{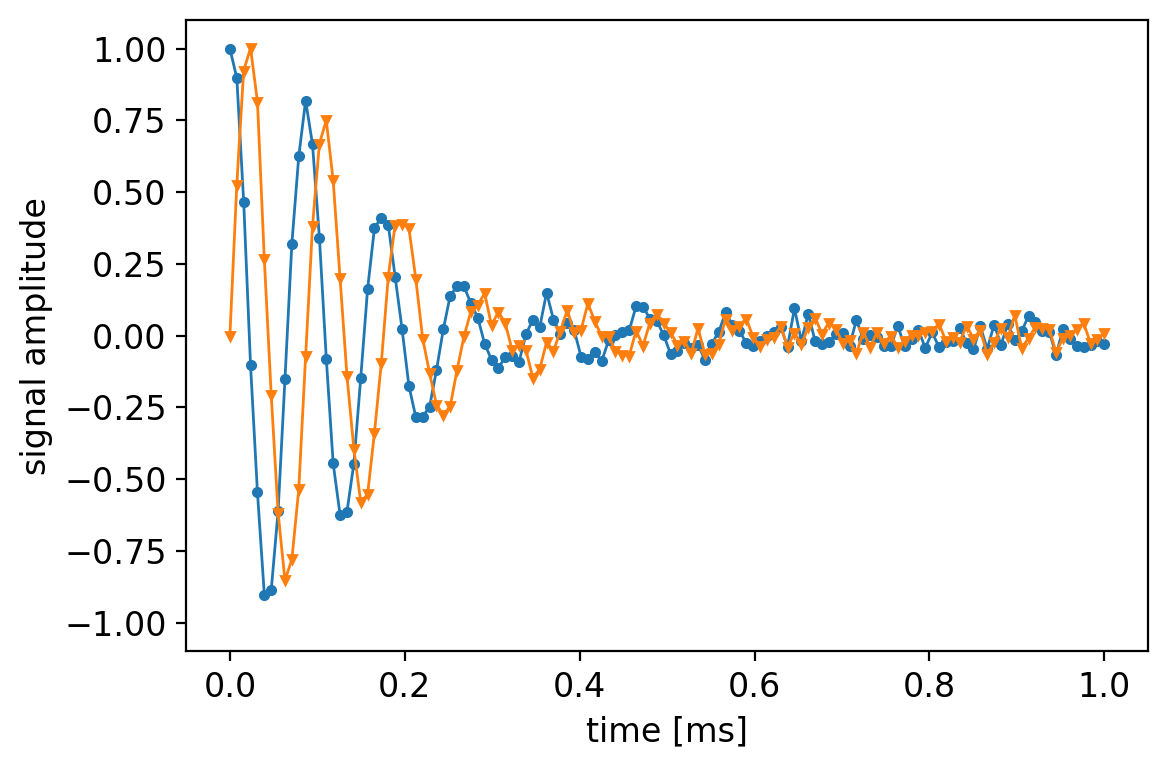}
    \caption{NMR signal, averaged over 1000~FID signals, with a total measurement time of roughly 7~seconds. Shown is the real (blue \ding{108}) and the imaginary part (orange \ding{116}) of the signal as a function of time. The signal amplitude is normalized to one at $t=0$~s. The exponential decay time of the signal is roughly 155~$\mu$s. The solid lines serve only as a guide for the eyes. }
    \label{fig:nmrSignal}
\end{figure}
Figure~\ref{fig:nmrSignal} shows an FID signal with a typical spectral width of 128~kHz, resulting in a sampling interval of 7.8125~$\mu$s. The signal is oscillating at 12.5~kHz which represents the difference between the pulse frequency of 21.68~MHz and the actual precession frequency of the protons. The reason for this is that the signal from the spin precession is digitally mixed with the reference signal from the oscillator that generates the pulse. The mixer produces an output that is the sum and the difference of the two signals. After a low-pass filter, only the difference signal remains. To get the real and imaginary part of the signal, the NMR system has actually two digital mixers. The spin precession signal is fed into both but the reference signal is phase-shifted by $\pi/2$ before one of the mixers. This allows quadrature detection to distinguish positive and negative frequencies without employing two pick-up coils. A detailed treatment on NMR techniques can be found in reference~\cite{keeler_understanding_2002}.

\subsection{Analysis Tools}\label{sec:analysisTools}

The precession signal of the FID is usually analyzed in the frequency domain after performing a fast Fourier transform (FFT)~\cite{heideman_gauss_1985}. Standard tools in signal processing are zero-padding and the application of window functions. The former adds zeros at the end of the signal in the time domain. This results in a higher spectral resolution. It does not add information but rather interpolates between points that are already there and makes the spectrum look smoother. The latter damps parts of the signal with a lower signal amplitude which reduces the noise in the spectrum. In the case of an FID, usually an exponential window is applied. We do not use either of them in our signal processing but fit the spectrum with a Lorentzian function
\begin{equation}\label{eq:lorentzFunction}
    S w \times \left( \frac{1}{1 + (f - f_0)^2 \ w^2} - i \frac{(f-f_0) \ w}{1 + (f-f_0)^2 \ w^2} \right) \times e^{-i \phi} + o \ ,
\end{equation}
where $S$ is the signal amplitude, $f_0$ the resonance frequency, $o$ the offset, and $w$ the scale parameter that is connected to the observed transversal relaxation time constant via ${T_2^* = w / 2 \pi}$. The real and imaginary parts correspond to the absorption and dispersion modes, respectively.~\footnote{This is the common definition in the NMR community~\cite{slichter_principles_1978,keeler_understanding_2002}. } The two can be mixed with the phase $\phi$. Ideally, the phase $\phi$ is zero and the absorption and dispersion modes of the Lorentzian are not mixed. Drifts in the NMR electronics can lead to a change of $\phi$. This changes the apparent spectral amplitude if only the absorption mode is considered, which leads to a systematic error. There are various methods to detect and correct this phase~ \cite{craig_automated_1988,chen_efficient_2002,bao_robust_2013}. We include the signal phase in the parameters of the fit to avoid this problem.  This makes the amplitude independent of the phase. 

\begin{figure}[!t]
    \centering
    \includegraphics[width=0.48\textwidth]{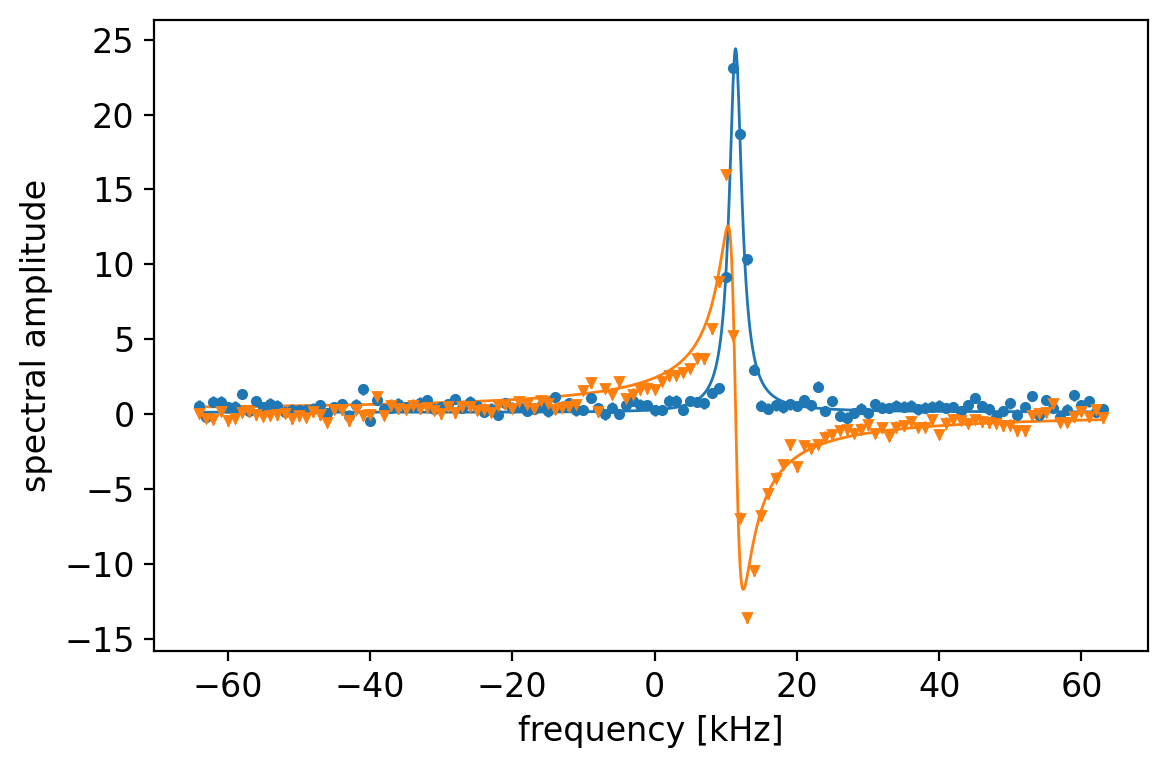}
    \caption{Spectrum of the NMR signal of  Fig.~\ref{fig:nmrSignal} that was calculated using an FFT algorithm. The spectral peak has the shape of a Lorentzian where the real part (blue \ding{108}) is the absorption mode and the imaginary part (orange \ding{116}) the dispersion mode. The solid lines are a least-squares fit of Eq.~(\ref{eq:lorentzFunction}). The absorption peak has a full width at half maximum (FWHM) of roughly 2.2~kHz. }
    \label{fig:nmrSpectrum}
\end{figure}

The noise of the spectrum can be estimated using the data points in the baseline and calculating their standard deviation around the mean. This uncertainty can then be provided to the fitting routine, which includes it in the calculation of the errors of the fit parameters. The NMR spectrum of the signal of Fig.~\ref{fig:nmrSignal} with a fitted Lorentzian of Eq.~(\ref{eq:lorentzFunction}) is shown in Fig.~\ref{fig:nmrSpectrum}. The fit yields $S = (26.2 \pm 0.3)$~kHz, $f_0 = (11.37 \pm 0.02)$~kHz, $w = (927 \pm 15)$~$\mu$s, $\phi = (-1.4 \pm 0.6)\degree$, and $o = (0.13 \pm 0.02)$ with $\chi^2 = 1058$ for 251 degrees of freedom. The resulting value for the reduced $\chi^2$ of about 4.2 is bigger than one because the fitting function Eq.~(\ref{eq:lorentzFunction}) does not account for all the characteristics of the NMR signal. The apparent amplitude in Fig.~\ref{fig:nmrSpectrum} is $Sw = 24.3 \pm 0.5$. From the fit value of $w$, the observed transversal relaxation time constant can be calculated, resulting in $T_2^* = (148 \pm 2)~\mu$s. This is comparable to the exponential decay time of the FID signal of $(155 \pm 7)~\mu$s.  

Depending on the measurement performed in the interaction region, the spin polarization and therefore the NMR signal amplitude may be small. To improve the fit of the spectrum we take a reference measurement before each measurement sequence. This measurement has the same parameters for the NMR pulse sequence, but all spin-flip signals in the interaction region are turned off. This leads to a reference signal with the highest possible amplitude. The Lorentzian in Eq.~(\ref{eq:lorentzFunction}) is then fitted to the reference spectrum and the parameters $f_0$ and $\phi$ are fixed for the fits of the subsequent measurements. Additionally, the amplitude of the reference is used for normalization, resulting in the \textit{normalized spin polarization} used in the measurements presented in Sec.~\ref{sec:measurements}.

\section{Measurements}\label{sec:measurements}

To characterize the apparatus we conducted several resonance measurements. We performed Rabi measurements~\cite{rabi_molecular_1939, kellogg_magnetic_1939} with a single spin-flip coil in the interaction region. An oscillating current is applied to create the $B_1$ field which causes a spin flip if the frequency is close to the Larmor resonance frequency of the proton spins in the $B_0$ field. The $B_1$ field is linearly oscillating and orthogonal to $B_0$, in our case it is aligned with the direction of the water flow along
the $z$-axis. We investigated the resonance at various $B_0$ fields and flow velocities.
\begin{figure}[!t]
  \centering
  \begin{subfigure}[b]{0.48\textwidth}
    \includegraphics[width=\textwidth]{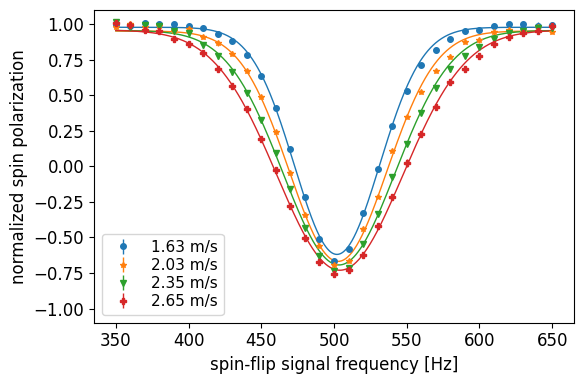}
    \caption{~}
    \label{fig:rabiFrequencyScan_a}
  \end{subfigure}
  \hfill
  \begin{subfigure}[b]{0.48\textwidth}
    \includegraphics[width=\textwidth]{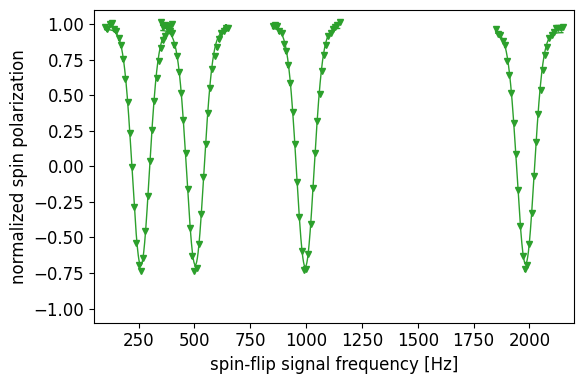}
    \caption{~}
    \label{fig:rabiFrequencyScan_b}
  \end{subfigure}
  \caption{ (a) Rabi resonances at $B_0 \approx 12~\mu$T ($\omega_0 \approx 2 \pi \times 500$~Hz) for water flow velocities of 1.63~m/s (blue \ding{108}), 2.03~m/s (orange \ding{72}), 2.35~m/s (green \ding{116}), and 2.65~m/s (red \ding{58}). The Gaussian fit of the data (solid lines) led to a FWHM of ($69.4 \pm 0.5$)~Hz, ($79.3 \pm 0.5$)~Hz, ($91.0 \pm 0.6$)~Hz, and ($101.9 \pm 0.6$)~Hz, respectively. The width scales proportionally to the velocity. (b) Rabi resonances at a water flow velocity of 2.35~m/s for various $B_0$ field settings, resulting in resonance frequencies at 250~Hz, 500~Hz, 1000~Hz, and 2000~Hz. More values of $B_0$ up to a resonance frequency of 10~kHz were explored but not shown here for reasons of legibility. }
  \label{fig:rabiFrequencyScan}
\end{figure} 
The current through the $B_0$-coil for the measurements shown in Fig.~\ref{fig:rabiFrequencyScan_a} was set to 50~mA. This creates a magnetic field of approximately 12~$\mu$T which corresponds to a resonance frequency of roughly 500~Hz. Note that the width of the resonance is proportional to the flow velocity. Additionally, the amplitude of the resonance increases since the time for spin relaxation is shorter. The amplitude of the spin-flip field $B_1$ used for these measurements was optimized for each velocity to achieve a $\pi$-flip on resonance.

For the measurements shown in Fig.~\ref{fig:rabiFrequencyScan_b}, the water flow velocity was fixed to 2.35~m/s. We measured the Rabi resonance for various values of the $B_0$ field resulting in resonance frequencies between 250~Hz and 10~kHz. Only four of the obtained resonance curves are shown. The full width at half maximum (FWHM) of $(90.9 \pm 0.2)$~Hz is identical for all resonances within the uncertainty. The measured width deviates from the theoretical value of 125~Hz when calculating the FWHM using the Rabi formula~\cite{piegsa_neutron_2015}. This can be explained by the fringe field of the spin-flip coil and leads to an effective length of the coil of 20~mm, compared to its geometric length of 15~mm. 

\begin{figure}[!t]
    \centering
    \includegraphics[width=0.48\textwidth]{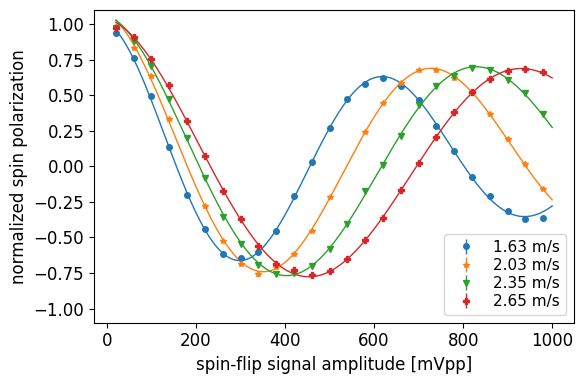}
    \caption{Rabi amplitude scan on resonance at $B_0 \approx 12~\mu$T ($\omega_0 \approx 2 \pi \times 500$~Hz) with the same velocities and colors as in Fig.~\ref{fig:rabiFrequencyScan_a}. The data are fitted with a sinusoidal function multiplied with an exponential decay (solid lines).} 
    \label{fig:rabiAmplitudeScan}
\end{figure}

To optimize the amplitude of the spin-flip signal for a $\pi$-flip, we scanned $B_1$ on resonance. A $\pi$-flip is achieved when the polarization is at the first minimum in Fig.~\ref{fig:rabiAmplitudeScan} at a value of the spin-flip signal amplitude between 300~mV$_\mathrm{pp}$ and 500~mV$_\mathrm{pp}$. The same measurement also demonstrates that a higher field amplitude is needed to achieve the same flipping angle if the water flows faster. 

\begin{figure}[!t]
  \centering
  \begin{subfigure}[b]{0.48\textwidth}
    \includegraphics[width=\textwidth]{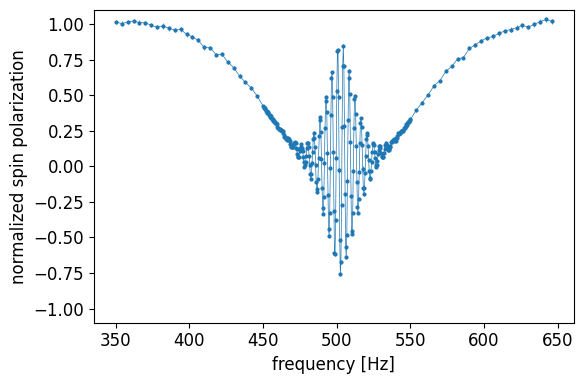}
    \caption{~}
  \end{subfigure}
  \hfill
  \begin{subfigure}[b]{0.48\textwidth}
    \includegraphics[width=\textwidth]{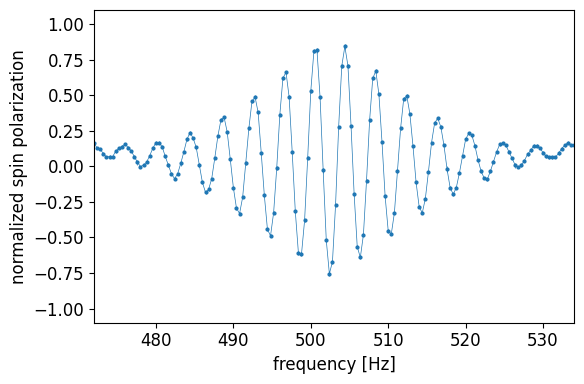}
    \caption{~}
  \end{subfigure}
  \caption{(a) Typical resonance pattern obtained with Ramsey's method of separated oscillatory fields at a magnetic field of $B_0 \approx 12$~$\mu$T. The average velocity of the water was 2.35~m/s. (b) Zoom in the central frequency range of the full resonance. The measurement time per point was roughly 18~s and the total measurement time 1.5~hours. The solid lines serve only as a guide for the eyes.}
  \label{fig:ramseyFrequencyScan}
\end{figure}

A typical measurement performed with two spin-flip coils in the interaction region is a Ramsey measurement~\cite{ramsey_new_1949,ramsey_molecular_1950}. In this type of measurement, the spins are first flipped by $\pi/2$. Then, they precess freely before they are flipped again by $\pi/2$. The two spin-flip signals are phase-locked and running at the same frequency. If this frequency is scanned over the resonance, a typical Ramsey pattern as shown in Fig.~\ref{fig:ramseyFrequencyScan} is obtained. As described in Sec.~\ref{sec:interactionRegion}, we placed the two spin-flip coils with a center-to-center separation of 600~mm. The overall envelope arises from the Rabi resonance. The fringes in the central region are the Ramsey interference pattern of the two spin-flip coils. The fringe period decreases with a longer distance between the spin-flip coils, making the experiment more sensitive. The visibility of the fringes above and below the resonance frequency is reduced due to the velocity distribution of the water in the capillary. In this measurement, the fringe period is approximately 4~Hz which is in good agreement with the expected value of $\frac{2.35~\mathrm{m/s}}{0.6~\mathrm{m}} \approx 3.9$~Hz.

\begin{figure}[!t]
  \centering
  \begin{subfigure}[b]{0.48\textwidth}
    \includegraphics[width=\textwidth]{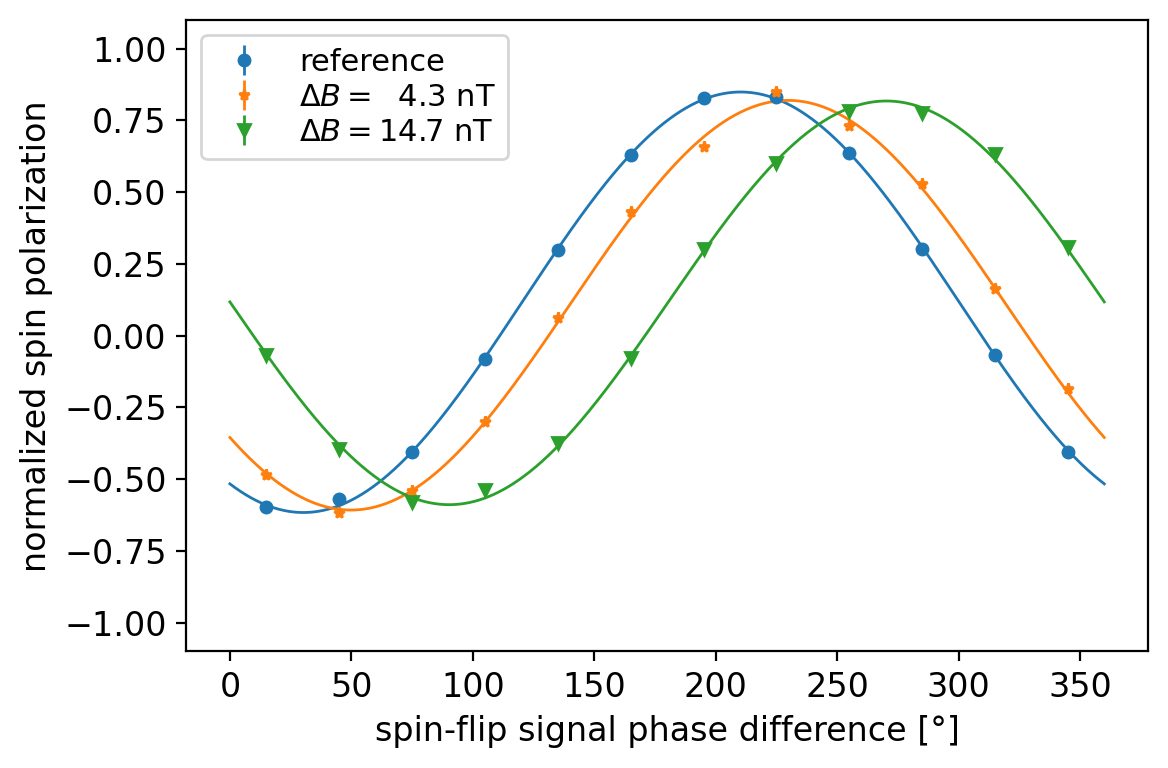}
    \caption{~}
    \label{fig:ramseyPhaseScan}
  \end{subfigure}
  \hfill
  \begin{subfigure}[b]{0.48\textwidth}
    \includegraphics[width=\textwidth]{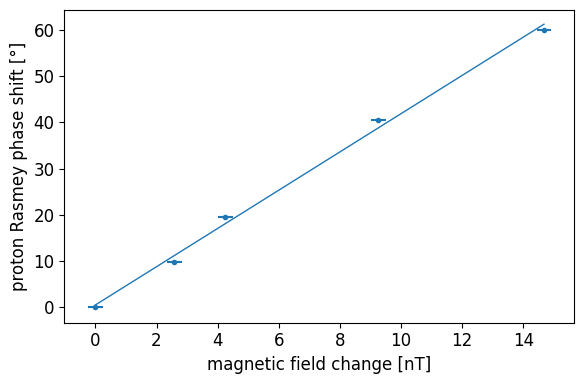}
    \caption{~}
    \label{fig:ramseyPhaseShift}
  \end{subfigure}
  \caption{(a) Ramsey phase scans on resonance at $\omega_0 \approx 2 \pi \times 500$~Hz, corresponding to a magnetic field of $B_0 \approx 12$~$\mu$T. Shown is the normalized spin polarization as a function of the phase between the two spin-flip signals. The water flow velocity was 2.35~m/s. To investigate the response of the setup to magnetic field changes, we performed a reference measurement (blue \ding{108}) and changed the magnetic field $B_0$ slightly to see the change in phase (orange \ding{72} and green \ding{116}). The data of each phase scan are fitted with a sinusoidal function with a fixed period of 360\degree (solid lines). The measurement time per phase scan is roughly 2~min. (b) Measured phase shift $\Delta \varphi$ as a function of the magnetic field change $\Delta B$. The linear fit (solid line) results in a slope of $\Delta \varphi / \Delta B = (4.1 \pm 0.1)$\degree/nT.}
  \label{fig:ramseyPhase}
\end{figure}

Another option for a Ramsey-type measurement is to keep the frequencies of both spin-flip signals fixed on resonance but instead scan the phase difference between the two oscillating signals. This has the advantage of always being on resonance, resulting in a measurement free from any frequency dispersion effects. Moreover, the obtained data are in the shape of a simple sinusoidal curve that can be fitted easily. The signals of such phase scans are shown in Fig.~\ref{fig:ramseyPhaseScan}. 
Ramsey's technique is very sensitive to (pseudo-)magnetic field effects. A tiny change in the magnetic field $\Delta B$ results in a change of the phase that the proton spins acquire in the interaction region
\begin{equation}\label{eq:ramseyPhase}
    \Delta \varphi = \gamma_p \Delta B \frac{L_\mathrm{eff}}{v} \ ,
\end{equation}
where $v$ is the water flow velocity and $L_\mathrm{eff}$ the effective interaction length which is slightly longer than the center-to-center separation of the spin-flip coils.~\footnote{The reason is, that the spins start already to precess within the spin-flip coil when partially flipped~\cite{piegsa_ramsey_2008}.} Such a shift of the Ramsey phase is visible in Fig.~\ref{fig:ramseyPhaseScan} where we changed the magnetic field by ($4.3 \pm 0.3$)~nT and ($14.7 \pm 0.3$)~nT. Note, these values were measured with the fluxgate after the second spin-flip coil and the average field between the spin-flip coils might be slightly different. These values of $\Delta B$ resulted in a phase shift of $19.5 \degree \pm 0.4 \degree$ and $60.0 \degree \pm 0.4 \degree$, respectively. 
To get the phase shift as a function of the magnetic field change we performed several additional measurements. A linear fit through the data results in a value of $\Delta \varphi / \Delta B = (4.1 \pm 0.1)$\degree/nT as presented in Fig.~\ref{fig:ramseyPhaseShift}. This agrees, within the limits of uncertainty, with the theoretical value that can be calculated using Eq.~(\ref{eq:ramseyPhase}). 

\begin{figure}[!t]
  \centering
  \begin{subfigure}[b]{0.48\textwidth}
    \includegraphics[width=\textwidth]{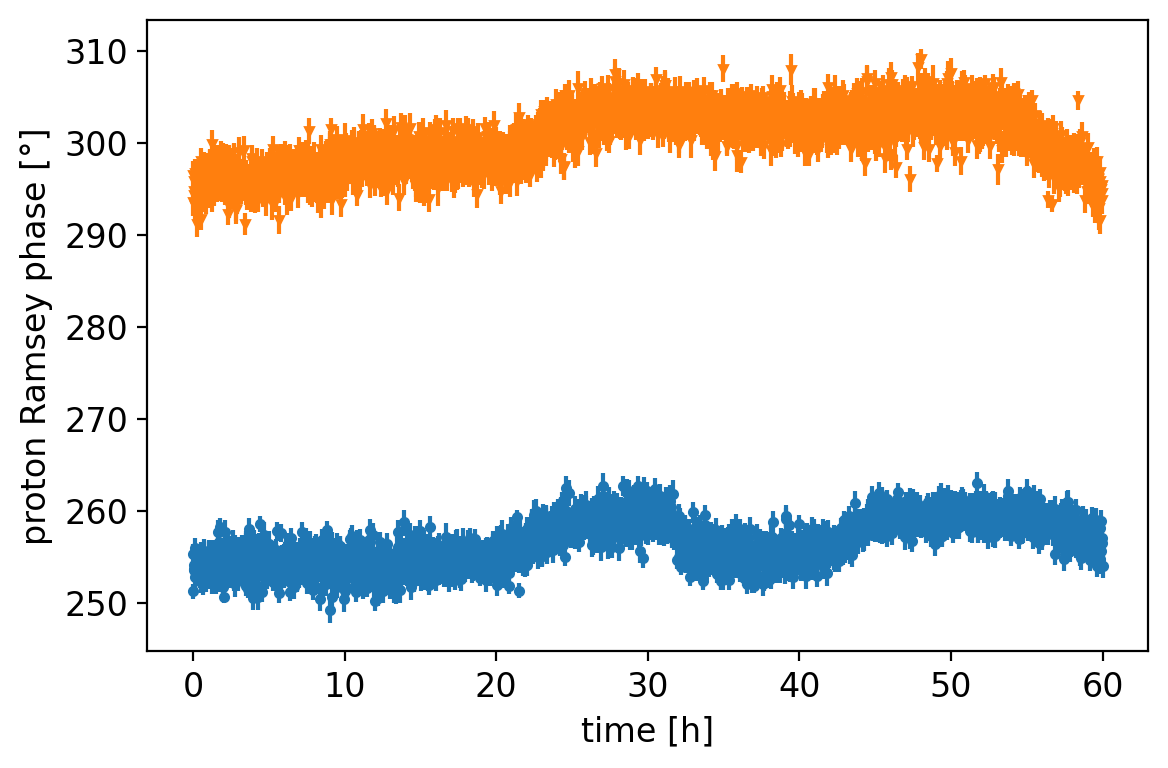}
    \caption{~}
    \label{fig:ramseyPhaseStability_timeSeries}
  \end{subfigure}
  \hfill
  \begin{subfigure}[b]{0.48\textwidth}
    \includegraphics[width=\textwidth]{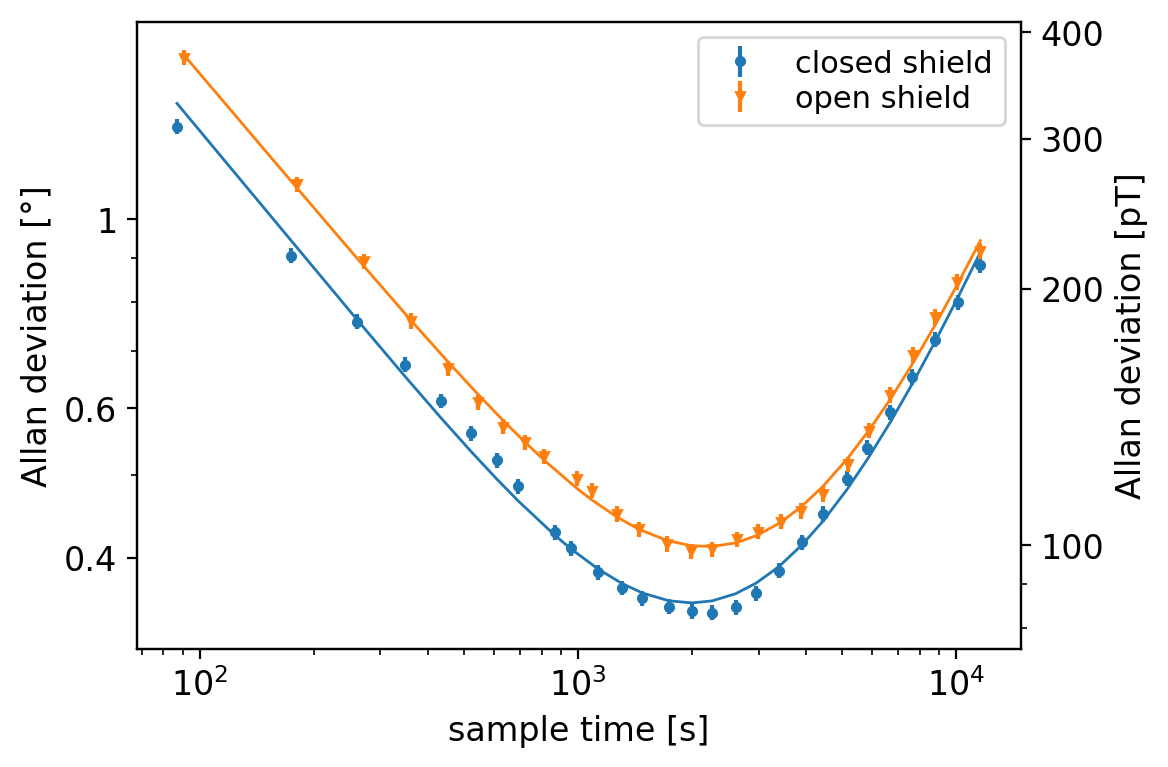}
    \caption{~}
    \label{fig:ramseyPhaseStability_allanDeviation}
  \end{subfigure}
  \caption{Proton phase stability for a setup with the end caps of the mu-metal shield installed (blue \ding{108}) and removed (orange \ding{116}). (a) Time series of the two consecutive measurements shown on the same time axis. (b) Overlapping Allan deviation of the phase as a function of the sample time. The vertical axis on the right side shows the corresponding magnetic field stability using the conversion factor of the linear fit in Fig.~\ref{fig:ramseyPhaseShift}. }
  \label{fig:ramseyPhaseStability}
\end{figure}

The sensitivity of the full apparatus is defined by the precision of the phase retrieval and the stability of the main magnetic field $B_0$. To characterize the stability, we performed phase scans as shown in Fig.~\ref{fig:ramseyPhaseScan} over 60~hours with the end caps of the mu-metal shield installed and removed in two consecutive measurement sequences. The stability was then analyzed by calculating the overlapping Allan deviation~\cite{allan_statistics_1966,howe_properties_1981} shown in Fig.~\ref{fig:ramseyPhaseStability}. 
The Allan deviations are fitted with
\begin{equation}
    \sigma_\mathrm{ASD} = \frac{a}{\sqrt{\tau}} + b \tau + c \ ,
\end{equation}
where $\tau$ is the sample time and $c$ is a constant offset. This fitting function is derived from reference~\cite{riley_handbook_2008}. The first term with coefficient $a$ dominates for short sample times and describes the statistical improvement of the uncertainty due to the averaging of the white noise. It approximately corresponds to the uncertainty at a sample time of 1~second and is $a = (13.7 \pm 0.2)\mathrm{\degree \, s^{1/2}}$ and $a = (15.6 \pm 0.2)\mathrm{\degree \, s^{1/2}}$ for the closed and the open shield, respectively. On the other hand, the second term with coefficient $b$ dominates for long sample times and describes the drifts of the system, i.e., magnetic field drifts. It is $b = (7.7 \pm 0.1) \times 10^{-5}~\mathrm{\degree \, s^{-1}}$ for both measurements. The constant offset was fitted to $c = -0.106\degree \pm 0.007\degree$ and $c = -0.089\degree \pm 0.009\degree$ for the closed and open shield, respectively. 
Both data sets show a minimum in the Allan standard deviation at about 30~minutes. The open shield has a minimum of $0.41\degree$ (98~pT) and the closed shield of $0.34\degree$ (83~pT). This implies, that after 30~minutes, a new reference measurement has to be taken.

\section{Conclusion}\label{sec:conclusion}

In conclusion, we presented a tabletop apparatus using proton spins in flowing water as probe particles. We used it in Rabi- and Ramsey-type measurements. We systematically investigated various parameter settings, e.g., the water flow velocity or the amplitude of the main magnetic field $B_0$. The setup works under a wide range of conditions and is sensitive to magnetic field effects below 100~pT. 
Our apparatus can be employed in a variety of fields and applications. In particular, we intend to use it in searches for new exotic long-range interactions beyond the standard model of particle physics.

\section*{Acknowledgments}
We gratefully acknowledge the excellent technical support by R.~H\"anni, S.~Bosco, J.~Christen, and L.~Meier from the University of Bern. This work was supported via the European Research Council under the ERC Grant Agreement no. 715031 (BEAM-EDM) and via the Swiss National Science Foundation under grants no. PP00P2-163663 and 200021-181996.

\section*{Conflict of Interest}
The authors have no conflicts to disclose.

\section*{Author Contributions}
\textbf{I.~Schulthess:} conceptualization (equal); data curation (lead); formal analysis (lead); investigation (lead); methodology (equal); software (lead); validation (equal); visualization (lead); writing - original draft (lead); writing – review \& editing (equal). 
\textbf{A.~Fratangelo:} writing – review \& editing (supporting). 
\textbf{P.~Hautle:} supervision (supporting); writing – review \& editing (supporting). 
\textbf{P.~Heil:} writing – review \& editing (supporting). 
\textbf{G.~Markaj:} writing – review \& editing (supporting). 
\textbf{M.~Persoz:} writing – review \& editing (supporting). 
\textbf{C.~Pistillo:} writing – review \& editing (supporting). 
\textbf{J.~Thorne:} writing – review \& editing (supporting). 
\textbf{F.~M.~Piegsa:} conceptualization (equal); funding acquisition (lead); methodology (equal); project administration (lead); resources (equal); supervision (lead); validation (equal); writing – review \& editing (equal).

\section*{Data Availability}
The data and analyis that support the findings of this study are openly available in a Github repository~\cite{schulthess_ivoschulthessprotonnmr_apparatus_2023}.


 \bibliographystyle{elsarticle-num} 
 \bibliography{references}

\begin{thebibliography}{10}
\expandafter\ifx\csname url\endcsname\relax
  \def\url#1{\texttt{#1}}\fi
\expandafter\ifx\csname urlprefix\endcsname\relax\def\urlprefix{URL }\fi
\expandafter\ifx\csname href\endcsname\relax
  \def\href#1#2{#2} \def\path#1{#1}\fi

\bibitem{rabi_molecular_1939}
I.~I. Rabi, S.~Millman, P.~Kusch, J.~R. Zacharias,
  \href{https://link.aps.org/doi/10.1103/PhysRev.55.526}{The {Molecular} {Beam}
  {Resonance} {Method} for {Measuring} {Nuclear} {Magnetic} {Moments}. {The}
  {Magnetic} {Moments} of \$\_3\${Li}\${\textasciicircum}6\$,
  \$\_3\${Li}\${\textasciicircum}7\$ and
  \$\_9\${F}\${\textasciicircum}\{19\}\$}, Physical Review 55~(6) (1939)
  526--535.
\newblock \href {https://doi.org/10.1103/PhysRev.55.526}
  {\path{doi:10.1103/PhysRev.55.526}}.
\newline\urlprefix\url{https://link.aps.org/doi/10.1103/PhysRev.55.526}

\bibitem{kellogg_magnetic_1939}
J.~M.~B. Kellogg, I.~I. Rabi, N.~F. Ramsey, J.~R. Zacharias,
  \href{https://link.aps.org/doi/10.1103/PhysRev.56.728}{The {Magnetic}
  {Moments} of the {Proton} and the {Deuteron}. {The} {Radiofrequency}
  {Spectrum} of {H}\$\_2\$ in {Various} {Magnetic} {Fields}}, Physical Review
  56~(8) (1939) 728--743.
\newblock \href {https://doi.org/10.1103/PhysRev.56.728}
  {\path{doi:10.1103/PhysRev.56.728}}.
\newline\urlprefix\url{https://link.aps.org/doi/10.1103/PhysRev.56.728}

\bibitem{ramsey_new_1949}
N.~F. Ramsey, \href{https://link.aps.org/doi/10.1103/PhysRev.76.996}{A {New}
  {Molecular} {Beam} {Resonance} {Method}}, Physical Review 76~(7) (1949)
  996--996.
\newblock \href {https://doi.org/10.1103/PhysRev.76.996}
  {\path{doi:10.1103/PhysRev.76.996}}.
\newline\urlprefix\url{https://link.aps.org/doi/10.1103/PhysRev.76.996}

\bibitem{ramsey_molecular_1950}
N.~F. Ramsey, \href{https://link.aps.org/doi/10.1103/PhysRev.78.695}{A
  {Molecular} {Beam} {Resonance} {Method} with {Separated} {Oscillating}
  {Fields}}, Physical Review 78~(6) (1950) 695--699.
\newblock \href {https://doi.org/10.1103/PhysRev.78.695}
  {\path{doi:10.1103/PhysRev.78.695}}.
\newline\urlprefix\url{https://link.aps.org/doi/10.1103/PhysRev.78.695}

\bibitem{ramsey_neutron_1986}
N.~F. Ramsey,
  \href{https://linkinghub.elsevier.com/retrieve/pii/0378436386903268}{Neutron
  magnetic resonance experiments}, Physica B+C 137~(1-3) (1986) 223--229.
\newblock \href {https://doi.org/10.1016/0378-4363(86)90326-8}
  {\path{doi:10.1016/0378-4363(86)90326-8}}.
\newline\urlprefix\url{https://linkinghub.elsevier.com/retrieve/pii/0378436386903268}

\bibitem{essen_atomic_1955}
L.~Essen, J.~V.~L. Parry, \href{https://www.nature.com/articles/176280a0}{An
  {Atomic} {Standard} of {Frequency} and {Time} {Interval}: {A} {Cæsium}
  {Resonator}}, Nature 176~(4476) (1955) 280--282.
\newblock \href {https://doi.org/10.1038/176280a0}
  {\path{doi:10.1038/176280a0}}.
\newline\urlprefix\url{https://www.nature.com/articles/176280a0}

\bibitem{wynands_atomic_2005}
R.~Wynands, S.~Weyers,
  \href{https://iopscience.iop.org/article/10.1088/0026-1394/42/3/S08}{Atomic
  fountain clocks}, Metrologia 42~(3) (2005) S64--S79.
\newblock \href {https://doi.org/10.1088/0026-1394/42/3/S08}
  {\path{doi:10.1088/0026-1394/42/3/S08}}.
\newline\urlprefix\url{https://iopscience.iop.org/article/10.1088/0026-1394/42/3/S08}

\bibitem{rosi_precision_2014}
G.~Rosi, F.~Sorrentino, L.~Cacciapuoti, M.~Prevedelli, G.~M. Tino,
  \href{http://www.nature.com/articles/nature13433}{Precision measurement of
  the {Newtonian} gravitational constant using cold atoms}, Nature 510~(7506)
  (2014) 518--521.
\newblock \href {https://doi.org/10.1038/nature13433}
  {\path{doi:10.1038/nature13433}}.
\newline\urlprefix\url{http://www.nature.com/articles/nature13433}

\bibitem{abel_measurement_2020}
C.~Abel, S.~Afach, N.~J. Ayres, C.~A. Baker, G.~Ban, G.~Bison, K.~Bodek,
  V.~Bondar, M.~Burghoff, E.~Chanel, Z.~Chowdhuri, P.-J. Chiu, B.~Clement,
  C.~B. Crawford, M.~Daum, S.~Emmenegger, L.~Ferraris-Bouchez, M.~Fertl,
  P.~Flaux, B.~Franke, A.~Fratangelo, P.~Geltenbort, K.~Green, W.~C. Griffith,
  M.~van~der Grinten, Z.~D. Grujić, P.~Harris, L.~Hayen, W.~Heil, R.~Henneck,
  V.~Hélaine, N.~Hild, Z.~Hodge, M.~Horras, P.~Iaydjiev, S.~N. Ivanov,
  M.~Kasprzak, Y.~Kermaidic, K.~Kirch, A.~Knecht, P.~Knowles, H.-C. Koch, P.~A.
  Koss, S.~Komposch, A.~Kozela, A.~Kraft, J.~Krempel, M.~Kuźniak, B.~Lauss,
  T.~Lefort, Y.~Lemière, A.~Leredde, P.~Mohanmurthy, A.~Mtchedlishvili,
  M.~Musgrave, O.~Naviliat-Cuncic, D.~Pais, F.~M. Piegsa, E.~Pierre, G.~Pignol,
  C.~Plonka-Spehr, P.~N. Prashanth, G.~Quéméner, M.~Rawlik, D.~Rebreyend,
  I.~Rienäcker, D.~Ries, S.~Roccia, G.~Rogel, D.~Rozpedzik, A.~Schnabel,
  P.~Schmidt-Wellenburg, N.~Severijns, D.~Shiers, R.~Tavakoli~Dinani, J.~A.
  Thorne, R.~Virot, J.~Voigt, A.~Weis, E.~Wursten, G.~Wyszynski, J.~Zejma,
  J.~Zenner, G.~Zsigmond,
  \href{https://link.aps.org/doi/10.1103/PhysRevLett.124.081803}{Measurement of
  the {Permanent} {Electric} {Dipole} {Moment} of the {Neutron}}, Physical
  Review Letters 124~(8) (2020) 081803.
\newblock \href {https://doi.org/10.1103/PhysRevLett.124.081803}
  {\path{doi:10.1103/PhysRevLett.124.081803}}.
\newline\urlprefix\url{https://link.aps.org/doi/10.1103/PhysRevLett.124.081803}

\bibitem{piegsa_new_2013}
F.~M. Piegsa,
  \href{https://journals.aps.org/prc/abstract/10.1103/PhysRevC.88.045502}{New
  {Concept} for a {Neutron} {Electric} {Dipole} {Moment} {Search} using a
  {Pulsed} {Beam}}, Physical Review C 88~(4), arXiv: 1309.1959 (Oct. 2013).
\newblock \href {https://doi.org/10.1103/PhysRevC.88.045502}
  {\path{doi:10.1103/PhysRevC.88.045502}}.
\newline\urlprefix\url{https://journals.aps.org/prc/abstract/10.1103/PhysRevC.88.045502}

\bibitem{chupp_electric_2019}
T.~Chupp, P.~Fierlinger, M.~Ramsey-Musolf, J.~Singh,
  \href{https://link.aps.org/doi/10.1103/RevModPhys.91.015001}{Electric dipole
  moments of atoms, molecules, nuclei, and particles}, Reviews of Modern
  Physics 91~(1) (2019) 015001.
\newblock \href {https://doi.org/10.1103/RevModPhys.91.015001}
  {\path{doi:10.1103/RevModPhys.91.015001}}.
\newline\urlprefix\url{https://link.aps.org/doi/10.1103/RevModPhys.91.015001}

\bibitem{abel_search_2017}
C.~Abel, N.~J. Ayres, G.~Ban, G.~Bison, K.~Bodek, V.~Bondar, M.~Daum,
  M.~Fairbairn, V.~V. Flambaum, P.~Geltenbort, K.~Green, W.~C. Griffith,
  M.~van~der Grinten, Z.~D. Grujić, P.~G. Harris, N.~Hild, P.~Iaydjiev, S.~N.
  Ivanov, M.~Kasprzak, Y.~Kermaidic, K.~Kirch, H.-C. Koch, S.~Komposch, P.~A.
  Koss, A.~Kozela, J.~Krempel, B.~Lauss, T.~Lefort, Y.~Lemière, D.~J.~E.
  Marsh, P.~Mohanmurthy, A.~Mtchedlishvili, M.~Musgrave, F.~M. Piegsa,
  G.~Pignol, M.~Rawlik, D.~Rebreyend, D.~Ries, S.~Roccia, D.~Rozpędzik,
  P.~Schmidt-Wellenburg, N.~Severijns, D.~Shiers, Y.~V. Stadnik, A.~Weis,
  E.~Wursten, J.~Zejma, G.~Zsigmond,
  \href{https://link.aps.org/doi/10.1103/PhysRevX.7.041034}{Search for
  axion-like dark matter through nuclear spin precession in electric and
  magnetic fields}, Physical Review X 7~(4) (Nov. 2017).
\newblock \href {https://doi.org/10.1103/PhysRevX.7.041034}
  {\path{doi:10.1103/PhysRevX.7.041034}}.
\newline\urlprefix\url{https://link.aps.org/doi/10.1103/PhysRevX.7.041034}

\bibitem{schulthess_new_2022}
I.~Schulthess, E.~Chanel, A.~Fratangelo, A.~Gottstein, A.~Gsponer, Z.~Hodge,
  C.~Pistillo, D.~Ries, T.~Soldner, J.~Thorne, F.~M. Piegsa,
  \href{https://link.aps.org/doi/10.1103/PhysRevLett.129.191801}{New {Limit} on
  {Axionlike} {Dark} {Matter} {Using} {Cold} {Neutrons}}, Physical Review
  Letters 129~(19) (2022) 191801.
\newblock \href {https://doi.org/10.1103/PhysRevLett.129.191801}
  {\path{doi:10.1103/PhysRevLett.129.191801}}.
\newline\urlprefix\url{https://link.aps.org/doi/10.1103/PhysRevLett.129.191801}

\bibitem{piegsa_limits_2012}
F.~Piegsa, G.~Pignol, \href{http://arxiv.org/abs/1205.0340}{Limits on the
  {Axial} {Coupling} {Constant} of {New} {Light} {Bosons}}, Physical Review
  Letters 108~(18), arXiv: 1205.0340 (May 2012).
\newblock \href {https://doi.org/10.1103/PhysRevLett.108.181801}
  {\path{doi:10.1103/PhysRevLett.108.181801}}.
\newline\urlprefix\url{http://arxiv.org/abs/1205.0340}

\bibitem{greene_measurement_1979}
G.~L. Greene, N.~F. Ramsey, W.~Mampe, J.~M. Pendlebury, K.~Smith, W.~B. Dress,
  P.~D. Miller, P.~Perrin,
  \href{https://link.aps.org/doi/10.1103/PhysRevD.20.2139}{Measurement of the
  neutron magnetic moment}, Physical Review D 20~(9) (1979) 2139--2153.
\newblock \href {https://doi.org/10.1103/PhysRevD.20.2139}
  {\path{doi:10.1103/PhysRevD.20.2139}}.
\newline\urlprefix\url{https://link.aps.org/doi/10.1103/PhysRevD.20.2139}

\bibitem{sherman_nuclear_1954}
C.~Sherman, \href{https://link.aps.org/doi/10.1103/PhysRev.93.1429}{Nuclear
  {Induction} with {Separate} {Regions} of {Excitation} and {Detection}},
  Physical Review 93~(6) (1954) 1429--1430.
\newblock \href {https://doi.org/10.1103/PhysRev.93.1429}
  {\path{doi:10.1103/PhysRev.93.1429}}.
\newline\urlprefix\url{https://link.aps.org/doi/10.1103/PhysRev.93.1429}

\bibitem{dobrescu_spin-dependent_2006}
B.~A. Dobrescu, I.~Mocioiu,
  \href{http://stacks.iop.org/1126-6708/2006/i=11/a=005?key=crossref.8f8f26ff813524b8579cb1f89a04a1c9}{Spin-dependent
  macroscopic forces from new particle exchange}, Journal of High Energy
  Physics 2006~(11) (2006) 005--005.
\newblock \href {https://doi.org/10.1088/1126-6708/2006/11/005}
  {\path{doi:10.1088/1126-6708/2006/11/005}}.
\newline\urlprefix\url{http://stacks.iop.org/1126-6708/2006/i=11/a=005?key=crossref.8f8f26ff813524b8579cb1f89a04a1c9}

\bibitem{schulthess_search_2022}
I.~Schulthess, \href{https://doi.org/10.48549/4103}{Search for {Axion}-{Like}
  {Dark} {Matter} and {Exotic} {Yukawa}-{Like} {Interaction}}, Ph.D. thesis,
  University of Bern (Dec. 2022).
\newline\urlprefix\url{https://doi.org/10.48549/4103}

\bibitem{bloch_magnetic_1940}
F.~Bloch, A.~Siegert,
  \href{https://link.aps.org/doi/10.1103/PhysRev.57.522}{Magnetic {Resonance}
  for {Nonrotating} {Fields}}, Physical Review 57~(6) (1940) 522--527.
\newblock \href {https://doi.org/10.1103/PhysRev.57.522}
  {\path{doi:10.1103/PhysRev.57.522}}.
\newline\urlprefix\url{https://link.aps.org/doi/10.1103/PhysRev.57.522}

\bibitem{cohen-tannoudji_absorption_1969}
C.~Cohen-Tannoudji, S.~Haroche,
  \href{http://www.edpsciences.org/10.1051/jphys:01969003002-3015300}{Absorption
  et diffusion de photons optiques par un atome en interaction avec des photons
  de radiofréquence}, Journal de Physique 30~(2-3) (1969) 153--168.
\newblock \href {https://doi.org/10.1051/jphys:01969003002-3015300}
  {\path{doi:10.1051/jphys:01969003002-3015300}}.
\newline\urlprefix\url{http://www.edpsciences.org/10.1051/jphys:01969003002-3015300}

\bibitem{muskat_dressed_1987}
E.~Muskat, D.~Dubbers, O.~Schärpf,
  \href{https://link.aps.org/doi/10.1103/PhysRevLett.58.2047}{Dressed
  {Neutrons}}, Physical Review Letters 58~(20) (1987) 2047--2050.
\newblock \href {https://doi.org/10.1103/PhysRevLett.58.2047}
  {\path{doi:10.1103/PhysRevLett.58.2047}}.
\newline\urlprefix\url{https://link.aps.org/doi/10.1103/PhysRevLett.58.2047}

\bibitem{tcs_micropumps_ltd_mgd2000_2021}
{TCS Micropumps ltd},
  \href{https://www.micropumps.co.uk/DATA/pdf/DS59%20-%20MGD2000%20Data%20Sheet%20REV%202.pdf}{{MGD2000}
  {Data} {Sheet}} (2021).
\newline\urlprefix\url{https://www.micropumps.co.uk/DATA/pdf/DS59%20-%20MGD2000%20Data%20Sheet%20REV%202.pdf}

\bibitem{thermo_fisher_scientific_thermo_2015}
{Thermo Fisher Scientific}, Thermo {Scientific} {Manual} (2015).

\bibitem{petley_temperature_1984}
B.~W. Petley, R.~W. Donaldson,
  \href{https://iopscience.iop.org/article/10.1088/0026-1394/20/3/002}{The
  {Temperature} {Dependence} of the {Diamagnetic} {Shielding} {Correction} for
  {Proton} {NMR} in {Water}}, Metrologia 20~(3) (1984) 81--83.
\newblock \href {https://doi.org/10.1088/0026-1394/20/3/002}
  {\path{doi:10.1088/0026-1394/20/3/002}}.
\newline\urlprefix\url{https://iopscience.iop.org/article/10.1088/0026-1394/20/3/002}

\bibitem{cho_nmr_2002}
H.~Cho, P.~B. Shepson, L.~A. Barrie, J.~P. Cowin, R.~Zaveri,
  \href{https://pubs.acs.org/doi/10.1021/jp020449%2B}{{NMR} {Investigation} of
  the {Quasi}-{Brine} {Layer} in {Ice}/{Brine} {Mixtures}}, The Journal of
  Physical Chemistry B 106~(43) (2002) 11226--11232.
\newblock \href {https://doi.org/10.1021/jp020449+}
  {\path{doi:10.1021/jp020449+}}.
\newline\urlprefix\url{https://pubs.acs.org/doi/10.1021/jp020449%2B}

\bibitem{supermagnete_datenblatt_2018}
{Supermagnete}, Datenblatt {Artikel} {Q}-40-10-05-{N} (Sep. 2018).

\bibitem{david_meeker_finite_2019}
{David Meeker}, \href{http://www.femm.info/wiki/HomePage}{Finite {Element}
  {Method} {Magnetics}} (2019).
\newline\urlprefix\url{http://www.femm.info/wiki/HomePage}

\bibitem{magnet-physik_dr_steingroever_gmbh_usb_2020}
{Magnet-Physik Dr. Steingroever GmbH},
  \href{https://www.magnet-physik.de/upload/38565932-HU-USB-Hall-Sonden-3167.pdf}{{USB}
  {Hall}-{Sonden}} (2020).
\newline\urlprefix\url{https://www.magnet-physik.de/upload/38565932-HU-USB-Hall-Sonden-3167.pdf}

\bibitem{bloch_nuclear_1946}
F.~Bloch, \href{https://link.aps.org/doi/10.1103/PhysRev.70.460}{Nuclear
  {Induction}}, Physical Review 70~(7-8) (1946) 460--474.
\newblock \href {https://doi.org/10.1103/PhysRev.70.460}
  {\path{doi:10.1103/PhysRev.70.460}}.
\newline\urlprefix\url{https://link.aps.org/doi/10.1103/PhysRev.70.460}

\bibitem{slichter_principles_1978}
C.~P. Slichter,
  \href{http://link.springer.com/10.1007/978-3-662-12784-1}{Principles of
  {Magnetic} {Resonance}}, Vol.~1 of Springer {Series} in {Solid}-{State}
  {Sciences}, Springer Berlin Heidelberg, Berlin, Heidelberg, 1978.
\newblock \href {https://doi.org/10.1007/978-3-662-12784-1}
  {\path{doi:10.1007/978-3-662-12784-1}}.
\newline\urlprefix\url{http://link.springer.com/10.1007/978-3-662-12784-1}

\bibitem{hahn_accurate_1949}
E.~L. Hahn, \href{https://link.aps.org/doi/10.1103/PhysRev.76.145}{An
  {Accurate} {Nuclear} {Magnetic} {Resonance} {Method} for {Measuring}
  {Spin}-{Lattice} {Relaxation} {Times}}, Physical Review 76~(1) (1949)
  145--146.
\newblock \href {https://doi.org/10.1103/PhysRev.76.145}
  {\path{doi:10.1103/PhysRev.76.145}}.
\newline\urlprefix\url{https://link.aps.org/doi/10.1103/PhysRev.76.145}

\bibitem{carr_effects_1954}
H.~Y. Carr, E.~M. Purcell,
  \href{https://link.aps.org/doi/10.1103/PhysRev.94.630}{Effects of {Diffusion}
  on {Free} {Precession} in {Nuclear} {Magnetic} {Resonance} {Experiments}},
  Physical Review 94~(3) (1954) 630--638.
\newblock \href {https://doi.org/10.1103/PhysRev.94.630}
  {\path{doi:10.1103/PhysRev.94.630}}.
\newline\urlprefix\url{https://link.aps.org/doi/10.1103/PhysRev.94.630}

\bibitem{meiboom_modified_1958}
S.~Meiboom, D.~Gill,
  \href{http://aip.scitation.org/doi/10.1063/1.1716296}{Modified
  {Spin}‐{Echo} {Method} for {Measuring} {Nuclear} {Relaxation} {Times}},
  Review of Scientific Instruments 29~(8) (1958) 688--691.
\newblock \href {https://doi.org/10.1063/1.1716296}
  {\path{doi:10.1063/1.1716296}}.
\newline\urlprefix\url{http://aip.scitation.org/doi/10.1063/1.1716296}

\bibitem{keysight_technologies_keysight_2020}
{Keysight Technologies},
  \href{https://www.keysight.com/us/en/support/B2962A/6-5-digit-low-noise-power-source-2-channels.html}{Keysight
  {B2961A}/{B2962A} 6.5 {Digit} {Low} {Noise} {Power} {Source}} (2020).
\newline\urlprefix\url{https://www.keysight.com/us/en/support/B2962A/6-5-digit-low-noise-power-source-2-channels.html}

\bibitem{keysight_technologies_keysight_2021}
{Keysight Technologies},
  \href{https://www.keysight.com/us/en/assets/7018-05928/data-sheets/5992-2572.pdf}{Keysight
  {33500B} and {33600A} {Series} {Trueform} {Waveform} {Generators}} (2021).
\newline\urlprefix\url{https://www.keysight.com/us/en/assets/7018-05928/data-sheets/5992-2572.pdf}

\bibitem{mini-circuits_coaxial_2011}
{Mini-Circuits},
  \href{https://www.minicircuits.com/pdfs/ZFRSC-2050+.pdf}{Coaxial {Power}
  {Splitter}/{Combiner} {ZFRSC}-2050+} (2011).
\newline\urlprefix\url{https://www.minicircuits.com/pdfs/ZFRSC-2050+.pdf}

\bibitem{pico_technology_tc-08_2021}
{Pico Technology},
  \href{https://www.picotech.com/download/datasheets/usb-tc-08-thermocouple-data-logger-data-sheet.pdf}{{TC}-08
  {Data} {Logger}} (2021).
\newline\urlprefix\url{https://www.picotech.com/download/datasheets/usb-tc-08-thermocouple-data-logger-data-sheet.pdf}

\bibitem{sensys_gmbh_sensys_2019}
{SENSYS GmbH}, \href{https://sensysmagnetometer.com/products/fgm3d/}{Sensys
  {FGM3D} {Datasheet}} (2019).
\newline\urlprefix\url{https://sensysmagnetometer.com/products/fgm3d/}

\bibitem{national_instruments_corporation_pcipxiusb-6289_2022}
{National Instruments Corporation}, {PCI}/{PXI}/{USB}-6289 {Specifications}
  (2022).

\bibitem{spincore_technologies_inc_ispin-nmr_2017}
{SpinCore Technologies, Inc.},
  \href{http://www.spincore.com/CD/iSpin/iSpin_Manual.pdf}{{iSpin}-{NMR}™
  {Owner}’s {Manual}} (2017).
\newline\urlprefix\url{http://www.spincore.com/CD/iSpin/iSpin_Manual.pdf}

\bibitem{spincore_technologies_inc_nmr_2022}
{SpinCore Technologies, Inc.},
  \href{http://www.spincore.com/products/Magnets/}{{NMR} {Permanent} {Magnets}}
  (2022).
\newline\urlprefix\url{http://www.spincore.com/products/Magnets/}

\bibitem{tiesinga_codata_2021}
E.~Tiesinga, P.~J. Mohr, D.~B. Newell, B.~N. Taylor,
  \href{https://link.aps.org/doi/10.1103/RevModPhys.93.025010}{{CODATA}
  recommended values of the fundamental physical constants: 2018}, Reviews of
  Modern Physics 93~(2) (2021) 025010.
\newblock \href {https://doi.org/10.1103/RevModPhys.93.025010}
  {\path{doi:10.1103/RevModPhys.93.025010}}.
\newline\urlprefix\url{https://link.aps.org/doi/10.1103/RevModPhys.93.025010}

\bibitem{bunting_magnetics_europe_bremag_2017}
{Bunting Magnetics Europe},
  \href{https://e-magnetsuk.com/wp-content/uploads/2020/06/E-Magnets-UK-Neodymium-Data-Sheet.pdf}{{BREMAG}
  {NdFeB} {Standard} {Range}} (2017).
\newline\urlprefix\url{https://e-magnetsuk.com/wp-content/uploads/2020/06/E-Magnets-UK-Neodymium-Data-Sheet.pdf}

\bibitem{kim_temperature_1998}
S.~Kim, C.~Doose,
  \href{http://ieeexplore.ieee.org/document/753163/}{Temperature compensation
  of {NdFeB} permanent magnets}, in: Proceedings of the 1997 {Particle}
  {Accelerator} {Conference} ({Cat}. {No}.{97CH36167}), Vol.~3, IEEE,
  Vancouver, BC, Canada, 1998, pp. 3227--3229.
\newblock \href {https://doi.org/10.1109/PAC.1997.753163}
  {\path{doi:10.1109/PAC.1997.753163}}.
\newline\urlprefix\url{http://ieeexplore.ieee.org/document/753163/}

\bibitem{hahn_spin_1950}
E.~L. Hahn, \href{https://link.aps.org/doi/10.1103/PhysRev.80.580}{Spin
  {Echoes}}, Physical Review 80~(4) (1950) 580--594.
\newblock \href {https://doi.org/10.1103/PhysRev.80.580}
  {\path{doi:10.1103/PhysRev.80.580}}.
\newline\urlprefix\url{https://link.aps.org/doi/10.1103/PhysRev.80.580}

\bibitem{keeler_understanding_2002}
J.~Keeler, Understanding {NMR} {Spectroscopy}, University of Cambridge,
  Department of Chemistry, 2002.

\bibitem{heideman_gauss_1985}
M.~T. Heideman, D.~H. Johnson, C.~S. Burrus,
  \href{http://link.springer.com/10.1007/BF00348431}{Gauss and the history of
  the fast {Fourier} transform}, Archive for History of Exact Sciences 34~(3)
  (1985) 265--277.
\newblock \href {https://doi.org/10.1007/BF00348431}
  {\path{doi:10.1007/BF00348431}}.
\newline\urlprefix\url{http://link.springer.com/10.1007/BF00348431}

\bibitem{craig_automated_1988}
E.~C. Craig, A.~G. Marshall,
  \href{https://linkinghub.elsevier.com/retrieve/pii/0022236488903502}{Automated
  phase correction of {FT} {NMR} spectra by means of phase measurement based on
  dispersion versus absorption relation ({DISPA})}, Journal of Magnetic
  Resonance (1969) 76~(3) (1988) 458--475.
\newblock \href {https://doi.org/10.1016/0022-2364(88)90350-2}
  {\path{doi:10.1016/0022-2364(88)90350-2}}.
\newline\urlprefix\url{https://linkinghub.elsevier.com/retrieve/pii/0022236488903502}

\bibitem{chen_efficient_2002}
L.~Chen, Z.~Weng, L.~Goh, M.~Garland,
  \href{http://linkinghub.elsevier.com/retrieve/pii/S1090780702000691}{An
  efficient algorithm for automatic phase correction of {NMR} spectra based on
  entropy minimization}, Journal of Magnetic Resonance 158~(1-2) (2002)
  164--168.
\newblock \href {https://doi.org/10.1016/S1090-7807(02)00069-1}
  {\path{doi:10.1016/S1090-7807(02)00069-1}}.
\newline\urlprefix\url{http://linkinghub.elsevier.com/retrieve/pii/S1090780702000691}

\bibitem{bao_robust_2013}
Q.~Bao, J.~Feng, L.~Chen, F.~Chen, Z.~Liu, B.~Jiang, C.~Liu,
  \href{https://linkinghub.elsevier.com/retrieve/pii/S109078071300150X}{A
  robust automatic phase correction method for signal dense spectra}, Journal
  of Magnetic Resonance 234 (2013) 82--89.
\newblock \href {https://doi.org/10.1016/j.jmr.2013.06.012}
  {\path{doi:10.1016/j.jmr.2013.06.012}}.
\newline\urlprefix\url{https://linkinghub.elsevier.com/retrieve/pii/S109078071300150X}

\bibitem{piegsa_neutron_2015}
F.~M. Piegsa,
  \href{https://linkinghub.elsevier.com/retrieve/pii/S0168900215003125}{A
  neutron resonance spin flip device for sub-millitesla magnetic fields},
  Nuclear Instruments and Methods in Physics Research Section A: Accelerators,
  Spectrometers, Detectors and Associated Equipment 786 (2015) 71--77.
\newblock \href {https://doi.org/10.1016/j.nima.2015.03.018}
  {\path{doi:10.1016/j.nima.2015.03.018}}.
\newline\urlprefix\url{https://linkinghub.elsevier.com/retrieve/pii/S0168900215003125}

\bibitem{piegsa_ramsey_2008}
F.~Piegsa, B.~van~den Brandt, H.~Glättli, P.~Hautle, J.~Kohlbrecher,
  J.~Konter, B.~Schlimme, O.~Zimmer,
  \href{https://linkinghub.elsevier.com/retrieve/pii/S0168900208002544}{A
  {Ramsey} apparatus for the measurement of the incoherent neutron scattering
  length of the deuteron}, Nuclear Instruments and Methods in Physics Research
  Section A: Accelerators, Spectrometers, Detectors and Associated Equipment
  589~(2) (2008) 318--329.
\newblock \href {https://doi.org/10.1016/j.nima.2008.02.020}
  {\path{doi:10.1016/j.nima.2008.02.020}}.
\newline\urlprefix\url{https://linkinghub.elsevier.com/retrieve/pii/S0168900208002544}

\bibitem{allan_statistics_1966}
D.~Allan, \href{http://ieeexplore.ieee.org/document/1446564/}{Statistics of
  atomic frequency standards}, Proceedings of the IEEE 54~(2) (1966) 221--230.
\newblock \href {https://doi.org/10.1109/PROC.1966.4634}
  {\path{doi:10.1109/PROC.1966.4634}}.
\newline\urlprefix\url{http://ieeexplore.ieee.org/document/1446564/}

\bibitem{howe_properties_1981}
D.~Howe, D.~Allan, J.~Barnes,
  \href{http://ieeexplore.ieee.org/document/1537481/}{Properties of {Signal}
  {Sources} and {Measurement} {Methods}}, in: Thirty {Fifth} {Annual}
  {Frequency} {Control} {Symposium}, IEEE, 1981, pp. 669--716.
\newblock \href {https://doi.org/10.1109/FREQ.1981.200541}
  {\path{doi:10.1109/FREQ.1981.200541}}.
\newline\urlprefix\url{http://ieeexplore.ieee.org/document/1537481/}

\bibitem{riley_handbook_2008}
W.~Riley, D.~Howe,
  \href{https://tsapps.nist.gov/publication/get_pdf.cfm?pub_id=50505}{Handbook
  of {Frequency} {Stability} {Analysis}}, Tech. Rep. 1065, Special Publication
  (NIST SP), National Institute of Standards and Technology, Gaithersburg, MD
  (2008).
\newline\urlprefix\url{https://tsapps.nist.gov/publication/get_pdf.cfm?pub_id=50505}

\bibitem{schulthess_ivoschulthessprotonnmr_apparatus_2023}
I.~Schulthess,
  \href{https://doi.org/10.5281/zenodo.7789189}{ivoschulthess/{protonNMR}\_apparatus:
  v1.0} (Mar. 2023).
\newblock \href {https://doi.org/10.5281/ZENODO.7789189}
  {\path{doi:10.5281/ZENODO.7789189}}.
\newline\urlprefix\url{https://doi.org/10.5281/zenodo.7789189}

\end{thebibliography}





\end{document}